%% file: main.tex
\documentclass[lettersize,journal]{IEEEtran}

\IEEEoverridecommandlockouts
\usepackage{amsmath,amssymb,amsfonts}
\usepackage{algorithmic}
\usepackage{graphicx}
\usepackage{textcomp}
\usepackage{xcolor}
\usepackage{enumitem}

\usepackage{import}
\import{}{imports}

\usepackage{cite}
\usepackage{etoolbox}

\def\BibTeX{{\rm B\kern-.05em{\sc i\kern-.025em b}\kern-.08em
    T\kern-.1667em\lower.7ex\hbox{E}\kern-.125emX}}
\begin{document}

\title{Exploiting Stragglers in Distributed Computing Systems with Task Grouping}

\author{Tharindu~Adikari,~\IEEEmembership{Graduate~Student~Member,~IEEE,}
Haider~Al-Lawati,~
Jason~Lam,~
Zhenhua~Hu,~
Stark~C.~Draper,~\IEEEmembership{Senior~Member,~IEEE}
\thanks{The work presented in this paper was performed while T. Adikari and H. Al-Lawati were affiliated with the University of Toronto. H. Al-Lawati is now with Oman Data Park, S. Draper is with the University of Toronto, and J. Lam and Z. Hu are with Huawei Technologies Canada.}
\thanks{This work was supported in part by Huawei Technologies Canada; and in part by the Natural Science and Engineering Research Council (NSERC) of Canada through a Discovery Research Grant.}
\thanks{This paper has been accepted for publication in IEEE Transactions on Services Computing. The initial results presented in this paper appeared in the proceedings of the Allerton Conference on Communication, Control, and Computing in 2023.}
}

\maketitle

\begin{abstract}
We consider the problem of stragglers in distributed computing systems. Stragglers, which are compute nodes that unpredictably slow down, often increase the completion times of tasks. One common approach to mitigating stragglers is work replication, where only the first completion among replicated tasks is accepted, discarding the others. However, discarding work leads to resource wastage. In this paper, we propose a method for exploiting the work completed by stragglers rather than discarding it. The idea is to increase the granularity of the assigned work, and to increase the frequency of worker updates. We show that the proposed method reduces the completion time of tasks via experiments performed on a simulated cluster as well as on Amazon EC2 with Apache Hadoop.
\end{abstract}


\input{body}
\section*{Acknowledgments}
The authors would like to thank Digital Research Alliance of Canada (formerly Compute Canada) for providing computing resources for some of the experiments.

\bibliographystyle{IEEEtran}
\bibliography{main}

\newcommand{\biospacing}{\vskip -22pt plus -1fil}

\begin{IEEEbiography}
[{\includegraphics[width=1in,height=1.25in,clip,keepaspectratio]{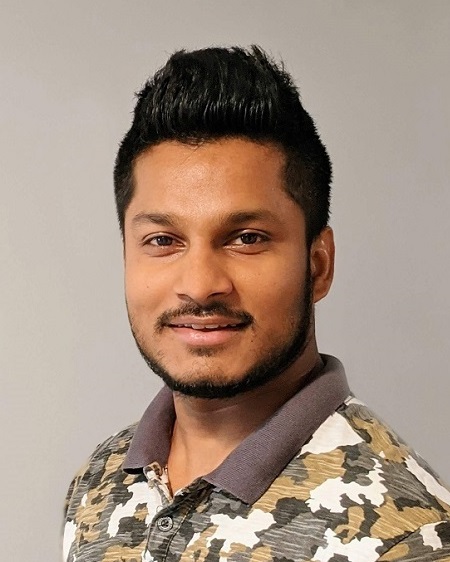}}]
{Tharindu Adikari} (Graduate Student Member, IEEE) obtained both PhD and MASc degrees from the Department of Electrical and Computer Engineering at the University of Toronto, Canada, in 2024 and 2018, respectively. He received the BSc degree in 2014 from the University of Moratuwa, Sri Lanka. From 2014 to 2016 he worked with LSEG Technology (formerly known as MillenniumIT), Sri Lanka, on stock market surveillance systems. His research interests include distributed computing, source coding, and machine learning.
\end{IEEEbiography}

\biospacing

\begin{IEEEbiography}
[{\includegraphics[width=1in,height=1.25in,clip,keepaspectratio]{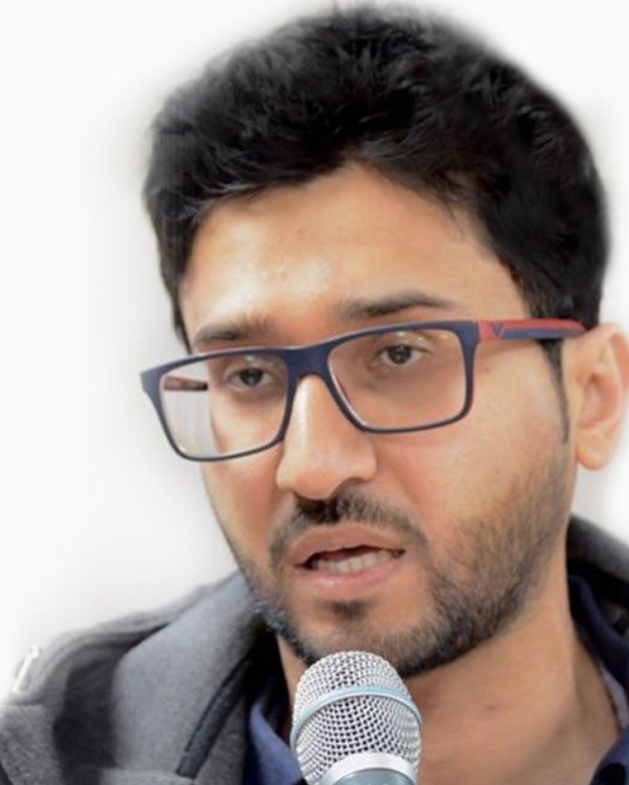}}]
{Haider Al-Lawati} received the B.Sc. and M.Sc. degrees in engineering and mathematics from Queen's University, Kingston, ON, Canada, in 2005 and 2007, respectively. He obtained the Ph.D. degree from the Electrical and Computer Engineering Department, University of Toronto. From 2008 to 2015, he worked with the Telecom Industry as an RF Planning Engineer, a Project Manager, and the Head of a Department working on various mobile technologies, such as GSM, WCDMA, LTE, and WiMAX. 
\end{IEEEbiography}

\biospacing

\begin{IEEEbiography}
[{\includegraphics[width=1in,height=1.25in,clip,keepaspectratio]{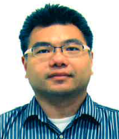}}]
{Jason Lam} is a Principle Distributed System Architect with Huawei Technologies Canada. He is a seasoned software engineer with background in distributed systems and big data; with primarily interests in system design, middleware and software optimization.
\end{IEEEbiography}

\biospacing

\begin{IEEEbiography}
[{\includegraphics[width=1in,height=1.25in,clip,keepaspectratio]{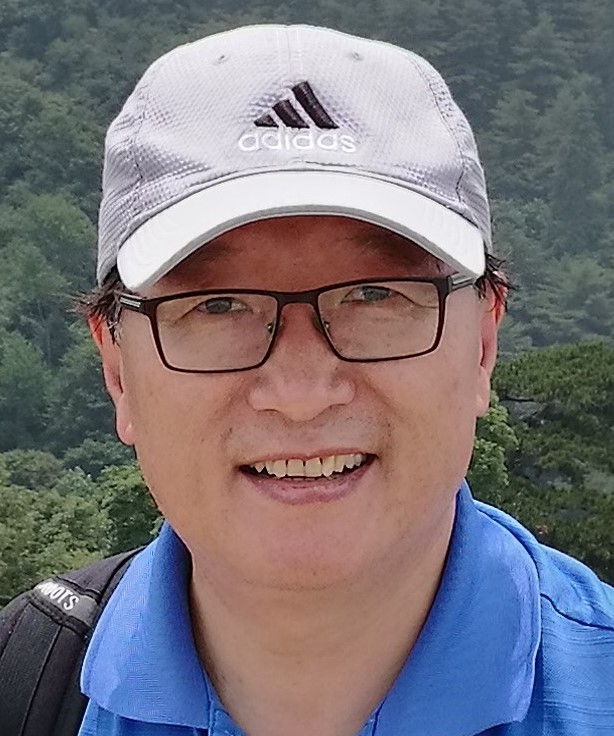}}]
{Zhenhua Hu} is a distinguished product architect and chief research architect in Huawei Technologies Canada. He also worked in IBM Canada as a senior principal product architect, and in Platform Computing as a principal product and research architect. His research interests include distributed computing, HPC, cloud computing, middleware, big data and artificial intelligence.
\end{IEEEbiography}

\biospacing

\begin{IEEEbiography}
[{\includegraphics[width=1in,height=1.25in,clip,keepaspectratio]{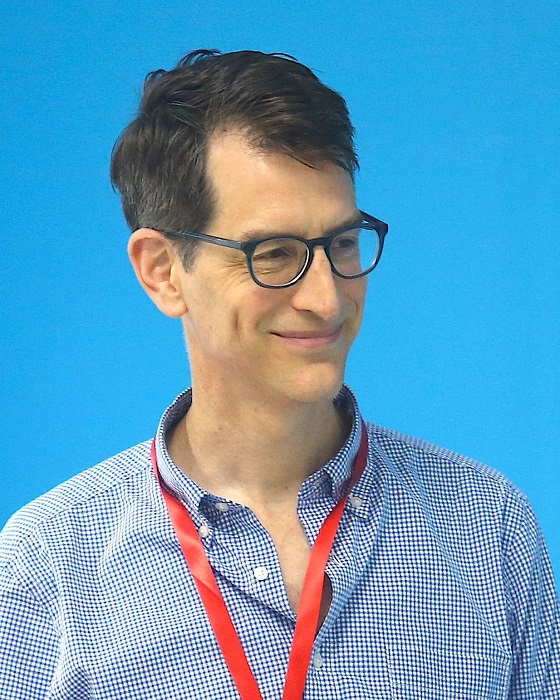}}]
{Stark C. Draper} (Senior Member, IEEE) received the B.S. degree in Electrical Engineering and the B.A. degree in History from Stanford University, and the M.S. and Ph.D. degrees in Electrical Engineering and Computer Science from the Massachusetts Institute of Technology (MIT).  He completed postdocs at the University of Toronto (UofT) and at the University of California, Berkeley. He is a Professor in the Department of Electrical and Computer Engineering at the University of Toronto and was an Associate Professor at the University of Wisconsin, Madison. As a Research Scientist he has worked at the Mitsubishi Electric Research Labs (MERL), Disney's Boston Research Lab, Arraycomm Inc., the C. S. Draper Laboratory, and Ktaadn Inc. His research interests include information theory, optimization, error-correction coding, security, and the application of tools and perspectives from these fields in communications, computing, learning, and astronomy. He has been the recipient of the NSERC Discovery Award, the NSF CAREER Award, the 2010 MERL President's Award, and teaching awards from UofT, the University of Wisconsin, and MIT. He received an Intel Graduate Fellowship, Stanford's Frederick E. Terman Engineering Scholastic Award, and a U.S. State Department Fulbright Fellowship. He spent the 2019–2020 academic year on sabbatical visiting the Chinese University of Hong Kong, Shenzhen, and the Canada-France-Hawaii Telescope (CFHT), Hawaii, USA. Among his service roles, he was the founding chair of the Machine Intelligence major at UofT, was the Faculty of Applied Science and Engineering (FASE) representative on the UofT Governing Council, is the FASE Vice-Dean of Research, and is the President of the IEEE Information Theory Society for 2024.
\end{IEEEbiography}

\end{document}

%% file: imports.tex
\usepackage{amsmath,amssymb}
\usepackage{mathtools}
\mathtoolsset{showonlyrefs,showmanualtags}
\usepackage{cancel}
\usepackage{graphicx}
\usepackage{bbm}
\usepackage[normalem]{ulem}
\usepackage{hyperref}
\usepackage{amsthm}
\usepackage[ruled]{algorithm2e}
\usepackage{multirow,array}
\usepackage{siunitx}
\usepackage{standalone}
\usepackage{pgfplots}
\usepackage{tikz}
\usepackage{subcaption}
\usepackage{import}
\import{}{notationCommon}

\import{}{comments}

\import{}{gww_chars}

%% file: notationCommon.tex
\usepackage{amsmath,amssymb}
\usepackage{mathtools}
\usepackage{etoolbox}


\makeatletter
\let\pragma@iinput=\@iinput
\def\@iinput#1{\xdef\@pragmafile{#1}\pragma@iinput{#1}}
\def\@pragmafile{default}
\def\pragmaonce{%
	\csname pragma@\@pragmafile\endcsname
	\global\expandafter\let \csname pragma@\@pragmafile\endcsname =  
}
\makeatother


\ifdefined\ceil
\else
\DeclarePairedDelimiter\ceil{\lceil}{\rceil}

\fi
\DeclarePairedDelimiterX{\inprd}[2]{\langle}{\rangle}{#1, #2}

\newcommand{\LnrmS}[1]{\left\lVert #1 \right\rVert}

\newcommand{\Rel}{\mathbb{R}}



\newcommand{\cc}[1]{\mathcal{#1}}

\newcommand{\calG}{\cc{G}}

\newcommand{\calL}{\cc{L}}
\newcommand{\calJ}{\cc{J}}


\definecolor{darkgreen}{rgb}{0.0, 0.5, 0.0}


\newcommand{\stkout}[1]{\ifmmode\text{\sout{\ensuremath{#1}}}\else\sout{#1}\fi}

\newcommand{\fig}[1]{{Fig.~\ref{fig:#1}}}

\newcommand{\secn}[1]{{Sec.~\ref{secn:#1}}}

\newcommand{\algo}[1]{{Alg.~\ref{algo:#1}}}

\newcommand{\eqn}[1]{{\eqref{eqn:#1}}}


\makeatletter
\newcommand*\ifcounter[1]{%
	\ifcsname c@#1\endcsname
	\expandafter\@firstoftwo
	\else
	\expandafter\@secondoftwo
	\fi
}
\makeatother

\ifcsmacro{theorem}{}{

	\ifcounter{chapter}{
		
	}{
		
	}
	
	
	\newcounter{example}[section]
	
	\newcounter{problem}[section]
	
}

\def\enumtheoremstart{\begin{enumerate}[noitemsep,label=(\roman*)]}
	\def\enumtheoremend{\end{enumerate}}

%% file: comments.tex
\usepackage{transparent}

\newif\ifshowcomments
\newif\ifshowdeleted

\newcommand{\devnull}[1]{}

\newcommand{\printcomment}[2]{\incolor{#1}{[[\ifx#1\empty\else#1: \fi#2]]}}

\newcommand{\ifequals}[3]{\ifthenelse{\equal{#1}{#2}}{#3}{}}
\newcommand{\case}[2]{#1 #2} 
\newenvironment{switch}[1]{\renewcommand{\case}{\ifequals{#1}}}{}

\definecolor{darkred}{rgb}{0.8, 0.01, 0.1}
\definecolor{darkgreen}{rgb}{0.0, 0.5, 0.0}
\definecolor{darkorange}{rgb}{0.93, 0.35, 0.1}

\newcommand{\incolor}[2]{\ignorespaces
	\begin{switch}{#1}\ignorespaces
		\case{TA}{\color{darkred}}\ignorespaces
		\case{SD}{\color{darkorange}}\ignorespaces
		\case{SK}{\color{darkgreen}}\ignorespaces
		\case{}{\color{red}}\ignorespaces
		#2
	\end{switch}
}

%% file: gww_chars.tex







\newcommand{\bx}{{\mathbf{x}}}

\newcommand{\by}{{\mathbf{y}}}





%% file: body.tex

\input{notationSpecific}

\newcommand{\reducebelowcaptionskip}{\setlength{\belowcaptionskip}{-10pt}}

\section{Introduction}\label{secn:backgroundamr}

\IEEEPARstart{A}n important challenge in distributed computing systems is the `straggler' problem caused by the variability of the processing power of compute nodes (which we also refer to as workers). In large compute clusters some workers may randomly or permanently become slower than others, delaying the completion of their assigned task. Such slow nodes are referred to as stragglers. The existence of stragglers in large compute clusters is well documented (e.g., see Fig. 1 in \cite{garraghan2016straggler}, Fig. 4 in \cite{wang2015using}, and Fig. 1(b) in \cite{ananthanarayanan2010reining}). The cause of stragglers may be due to multiple reasons, one of which is the low performance and unreliability of compute hardware. As commodity hardware has become cheaper and more affordable over time, it has been increasingly used to build compute clusters. The low performance and unreliability of such hardware worsen the straggler issue. Another cause of stragglers is the sharing of resources such as central processing unit (CPU), memory, and network \cite{dean2013tail}. The execution of multiple tasks on shared resources may lead to slowdowns which may give the impression of a random process. 

Task replication is a common straggler mitigation technique. In task replication, a task is replicated and multiple copies of the same task are assigned to different nodes. The result of only the first worker to finish the task is stored. Since some of the replications will be assigned to faster nodes, task replication can lead to smaller processing times. 
One important observation about task replication-based approaches for straggler mitigation is that the work done by the stragglers is discarded. 
In certain settings, e.g., \cite{ferdinand2018anytime, al-lawati2021esynchronous}, the authors propose methods for `exploiting' the work done by the stragglers rather than discarding it. 
But these proposals are developed for the special case of stochastic optimization algorithms (e.g., stochastic gradient descent and dual averaging). Such applications rely on the property that such algorithms are tolerant to some degree of approximate compute. 
Specifically, the authors propose to impose all workers a processing time limit. 
When this time limit is reached the workers stop computing and return the results. 
The advantage to this is that all workers finish their work at the same time (i.e., there is no variability in completion time), albeit some work may be only partially complete. Importantly, the work done by all workers is used, slow and fast workers alike. 
Theoretical guarantees show that such stochastic optimization algorithms perform well even when delivering approximate computations \cite{al-lawati2021esynchronous}. 
These results in stochastic optimization motivated us to explore application of the underlying principle and insights in other domains. 
In this paper, we show how the underlying ideas of straggler exploitation in \cite{al-lawati2021esynchronous} can be extended to non-stochastic optimization tasks.

\begin{figure}
\centering\includegraphics[width=0.9\columnwidth]{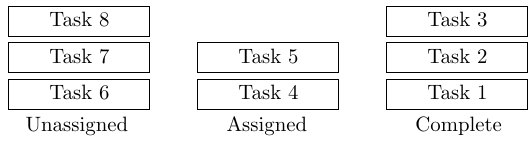}
\reducebelowcaptionskip
\caption{Visualizing task processing with $8$ tasks and $2$ workers. The two workers are processing task 4 and 5. If the `unassigned' stack is empty but the `assigned' stack is not, a worker will replicate the processing of a task in the `assigned' stack.}
\label{fig:task_stack}
\end{figure}

\fig{task_stack} presents the task processing model we consider. Here, a set of $8$ tasks are processed by $2$ workers. The three stacks show the `unassigned' tasks, i.e., the tasks that are yet to be processed, the `assigned' tasks, i.e., the tasks that are being processed by the two workers, and the `complete' tasks, i.e., the tasks that have been completely processed. 
The tasks are not necessarily of equal size. The two workers process tasks in the `unassigned' stack until the stack is empty. 
A `master' node may assist in centralized coordination of tasks and workers. For this type of tasks, a common straggler mitigation technique is task replication. When the `unassigned' stack is empty but the `assigned' stack is not, a worker starts processing a task in the `assigned' stack, thus replicating work. The result of the first worker to finish a task is stored. The results from other workers are discarded. This technique is useful in straggler-rich environments where some workers may take a long time to finish their tasks. Replicating work on a faster worker can lead to smaller processing times. In the sequel we refer to this type of replication as `standard' replication. 

The task processing model in \fig{task_stack} is standard in processing applications such as MapReduce \cite{dean2008mapreduce}, Spark \cite{zaharia2010spark} and Dryad \cite{isard2007dryad}. In this paper we experiment with the MapReduce framework, which is a widely used system that employs standard replication in its `map phase'. MapReduce is a popular programming framework for writing and executing parallel programs that process vast amounts of data on large clusters. In MapReduce, the implementation details common to many problems such as parallelization, fault-tolerance, and data distribution in the cluster are kept hidden from the user and are handled by the MapReduce framework. This abstraction enabled by the framework allows MapReduce to parallelize a task with minimal inputs from the user. In this paper we present some sample results obtained using MapReduce framework in straggler rich clusters.

\subsection{Our Contributions} 

Our contributions in this paper are two-fold. First, we introduce a method to exploit the work of stragglers based on generalizing standard replication. Second, we introduce a computational framework in which we can conduct repeatable experiments of a distributed system that suffers from stragglers. 
Our proposal for generalizing standard replication for exploiting work done by stragglers is as follows. We introduce a parameter we call `group size', denoted by $\groupSize$. Standard replication can be recovered by setting $\groupSize=1$. Our idea is to increase the granularity of tasks assigned to workers. We achieve this first by grouping tasks when assigning them to workers, and second by increasing the frequency of worker updates. These changes allow the proposed algorithm to make use of the progress made by slower workers rather than simply disregarding their work. 

The time savings that result from our algorithm come from two sources. The first is the reduction in overhead incurred when assigning work to a new worker. The second is the exploitation of stragglers. Grouping tasks is a well-known technique to obtain time savings by reducing overhead. We conduct experiments to tease out what fraction of the time reduction is due to reductions in overhead and how much is due to exploitation of stragglers. Experiments demonstrate that the time savings are due to both. 

As our second contribution we introduce a computational framework for conducting repeatable experiments with stragglers. 
In our framework we artificially induce stragglers to match a target straggler profile observed in a real system. The idea is first, to obtain a set of arbitrarily induced straggler profiles. We accomplish this by loading CPU cores using \texttt{stress-ng} tool, which artificially slows workers in a reproducible way. For each such straggler profile, we measure the completion time of a task and obtain completion time histogram. Second, given a target straggler histogram (obtained from a real system), we induce a convex combination of straggler profiles to match the target histogram. This is effectively a mixture model of stragglers. This method allows us to study different algorithms and parameter settings in a controlled manner where straggler characteristics are reproducible.

The initial results we present in this paper appeared in \cite{adikari2023straggler}. The two most important contributions we make in this paper are the presentation of the straggler simulating framework for conducting repeatable experiments, and the presentation of the real world experiments conducted on Amazon EC2, which we quickly summarize in the next section.

\subsection{Summary of Results}
We use two types of experiments to demonstrate that the completion time of tasks can be reduced by setting $\groupSize>1$. 
The first set of experiments are Monte-Carlo simulations. 
The second involves implementing the proposed algorithm using Apache Hadoop and executing on an Amazon EC2 cluster. 
While the Monte-Carlo experiments allow us to explore aspects of the proposed algorithm in a controlled manner, not possible in the EC2, the EC2 experiments provide real world verification and possible identification of unmodeled effects. 

For the Amazon EC2 experiments, we use two publicly available datasets: the Wikipedia dataset which comprises of textual data, and the LibriSpeech corpus which consists of audio recordings. With the Wikipedia dataset we conduct a series of text processing tasks, and with the LibriSpeech corpus we conduct a series of audio bit-rate conversion tasks. In our experiments we use Hadoop YARN as the resource management framework and Hadoop Distributed File System (HDFS) as the distributed file system. 
In our compute cluster each node consists of 8~cores, 32~GB memory and 100~GB EBS storage. 
More details regarding the experiment setups are provided in Sections~\ref{secn:simulateexp} and~\ref{secn:ec2exp}. 

\begin{figure}
	\reducebelowcaptionskip
	\centering\includegraphics[width=1\columnwidth]{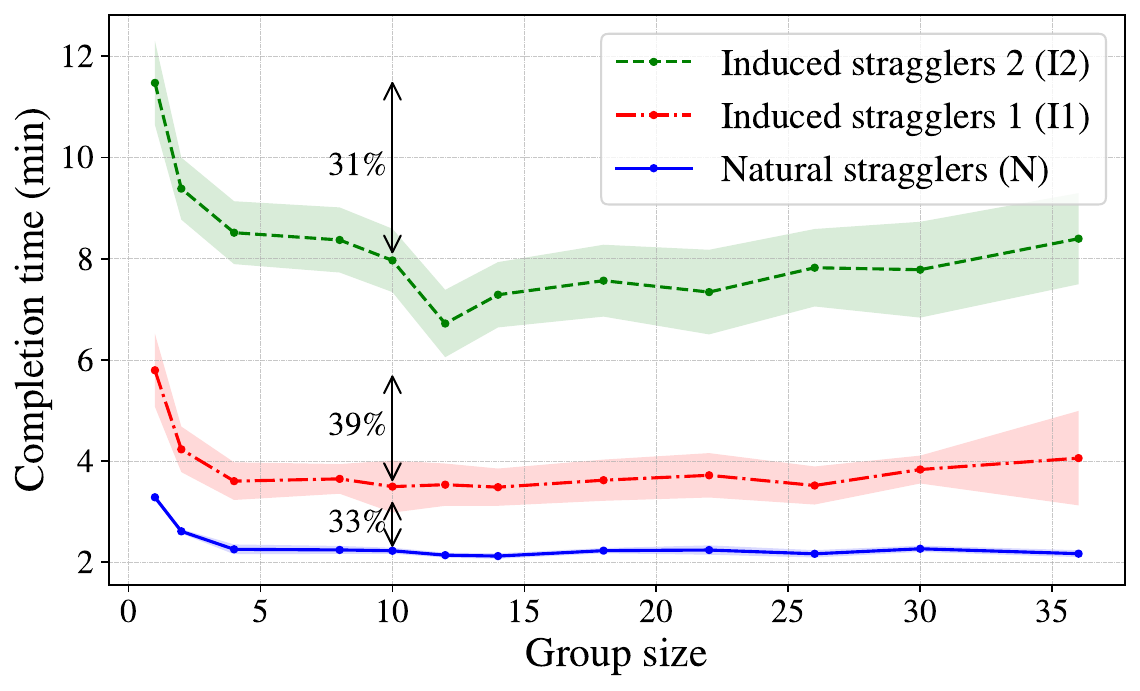}
	\caption{Completion time vs. group size parameter $\groupSize$. Plots are obtained by averaging 20 repetitions (shaded areas indicate standard deviations). Note that $\groupSize=1$ corresponds to standard replication. Setting $\groupSize>1$ yields gains on the order of $30-40\%$. In each plot the arrows show the percentage improvement at $\groupSize=10$ compared to $\groupSize=1$. Making $\groupSize$ too large slightly reduces the improvement.}
	\label{fig:yarntestresults}
\end{figure}

\fig{yarntestresults} presents sample results of how the completion time of the map phase of the algorithm varies with $\groupSize$ in our Amazon EC2 experiments with the Wikipedia dataset. As the labels suggest, the plots in \fig{yarntestresults} are obtained with two types of stragglers, `natural' and `artificially induced'. For the plot labeled N (`natural') we run our algorithm on the cluster as is, experimenting the natural processing time of the cluster. In contrast, for the plots labeled (I1) and (I2) we artificially induce stragglers on compute nodes to follow two different straggler statistics profiles. This helps us manipulate the cluster to mimic a straggler-rich operating environment. An important observation in \fig{yarntestresults} is that while the three plots achieve their minimums at slightly different values of $\groupSize$, setting $\groupSize=4$ is sufficient to gain most of the times savings in all cases. In all plots we observe that setting $\groupSize=10$ yields time savings of more than $30\%$ compared to $\groupSize=1$. Setting $\groupSize$ too large slightly increases completion time. We conjecture that this is because increasing $\groupSize$ also increases the read/write operations in HDFS that take place over the communication network, hence leading to network congestion leading to slow-down.

In addition to experiments involving compute clusters consisting of stragglers, we also conduct experiments with simulated `elastic' clusters, where workers are allowed to leave or join the cluster at short or no notice. The main advantage of such clusters is that service can be provided at a fraction of the cost of clusters with more stringent service level agreements. Services such as Amazon EC2 Spot and Microsoft Azure Batch are examples of cloud computing services that provide elastic clusters. Straggler mitigation algorithms exist \cite{dau2019optimizing, yang2019coded, kiani2021hierarchical} for elastic clusters when error correcting codes can be applied to the underlying tasks (e.g., matrix multiplication). Our experiments demonstrate that the proposed algorithm works well in exploiting stragglers in elastic clusters even with tasks that are not suitable for coding-based approaches.

The rest of this paper is organized as follows. 
In \secn{relatedworkmr} we briefly discuss existing methods for addressing the straggler problem. 
In \secn{proposedamr} we present the details of the proposed algorithm. 
In \secn{simulateexp} we present experiments conducted on the simulated cluster to better understand aspects and the characteristics of the scheme. 
In \secn{ec2exp} we provide implementation details of the proposed algorithm on Apache Hadoop, and we present the results of the experiments conducted on Amazon EC2. 
Finally, we present the concluding remarks in \secn{nextstepsmrf}.

\section{Related Work} \label{secn:relatedworkmr}

In standard replication, the workers replicate work when there are no more unassigned tasks. The result of the first worker to finish a replicated task is stored, and the results from other workers are discarded. The standard replication strategy can be improved by combining it with other techniques. One popular approach is to \emph{wait and speculate} the time it takes to complete tasks, so as to get a sense of how to prioritize task replication. This is done by measuring the `completion percentage' of tasks. Workers periodically report to master what percentage of the tasks assigned to them have been completed. Replicating a task being handled by a worker with a very low completion percentage may be more beneficial than replicating one being handled by a worker with a high completion percentage. 

Existing speculative methods differ in their implementations. For example, the LATE algorithm~\cite{zaharia2008improving} developed for MapReduce estimates the remaining time for completion of a task based on each task's completion percentage. This involves waiting and observing the periodic reports from workers before replicating a task. In addition to the information from periodic reporting, methods such as Mantri~\cite{ananthanarayanan2010reining} and MCP~\cite{chen2013improving} incorporate side information such as possible hardware failures in the cluster. Having some estimate of the relative performance of workers provides the master a sense of which tasks should be prioritized for replication first. Two examples are \cite{xu2013resource} and \cite{xu2014speculative} in which the authors present a theoretical framework for determining the task to be replicated. Another example is \cite{xu2015task} where authors propose a method for prioritizing tasks depending on the remaining processing time. In more recent work, the authors of \cite{zhou2018energy} propose an algorithm that improves performance by determining when and where (i.e., on which workers) to replicate work.

In contrast to wait and speculate methods that focus on prioritizing task replication, the Dolly algorithm introduced in \cite{ananthanarayanan2013effective} proposes full replication of a subset of the tasks at the very beginning. This algorithm assumes a distribution of task sizes and requires a priori knowledge of the sizes of the tasks. As per the experimental results presented in \cite{ananthanarayanan2013effective}, fully replicating all \emph{small} tasks offers large improvements in the completion time in the presence of stragglers. 

We note that our proposed generalization to standard replication is not intended to replace existing wait and speculate methods, but rather to complement them. In situations where task replication is necessary, our algorithm can be used in conjunction with an existing speculation-based method to prioritize the replication of tasks. While we anticipate even greater time savings through the application of speculation-based techniques in conjunction with our proposed method, we have opted not to explore this avenue in this paper as it would detract from our main message. 

Another straggler mitigation approach is to apply error correction coding techniques. This approach is well suited to tasks such as distributed matrix multiplication \cite{yu2017polynomial, dutta2019optimal, kiani2021hierarchical} and distributed stochastic gradient descent (SGD) \cite{tandon2017gradient, reisizadeh2019tree}, where the linear structure of the computation dovetails with the linear structure of the error correcting code. 
However, coding-based approaches may not be suitable to the task processing model we consider in this paper (cf. \fig{task_stack}). An example is when a non-linear function is to be applied to each image in a set of images (e.g., computing the histogram of an image). Although the task is easily parallelizable, it is not clear how to apply coding like in the matrix multiplication example. 

In other related work, the authors of \cite{amiri2019computation} propose and analyze a task scheduling algorithm in the context of statistical learning. In essence, the algorithm assigns all tasks to all workers, and tasks are executed in a predetermined order. Nevertheless, such an assignment might face challenges in compute clusters with limited storage and network bandwidth. 

There exists a large body of research on mitigating the load imbalance of workers when assigning workloads. These approaches primarily fall into two main categories: static scheduling algorithms, which are typically independent of the state of the system \cite{kokilavani2011load, daoud2008high, braun2001comparison}, and dynamic scheduling algorithms, which assign tasks depending on the current load of workers \cite{khan2024dynamic, mishra2017multi, ld2013honey, zhang1997comparison}. While intelligent scheduling of workloads reduces the runtime of task execution, that alone does not address issues arising due to random slowdowns of workers. 

Consider dynamic scheduling and our proposed algorithm. While both aim to minimize the time taken to complete all tasks, they address different problems. Dynamic scheduling assigns tasks to workers to minimize overloading. However, at the time of task assignment, one is unaware of the future straggler behaviour of the workers. Therefore, even with dynamic scheduling, replication might be needed to overcome the effects of random stragglers. Our algorithm addresses the question of how to make the most of the computations already completed by a worker that becomes a straggler. In summary, the proposed algorithm achieves this by increasing the granularity of the assigned work. While dynamic scheduling can be used to address load imbalance, our algorithm can be used on top of dynamic scheduling to address the random straggler issue. Therefore, the two approaches complement each other.

\section{Proposed Method} \label{secn:proposedamr}

The objective of the proposed method is to improve the completion time and to minimize wasted resources. Prior to discussing details, we list a few important parameters of the system. Let $\numWorkers$ and $\numSplits$ respectively denote the number of workers in the system and the number of tasks. In practice, there exists a time overhead for assigning work to a worker. This is due to allocation of hardware resources on workers and initializing them. The overhead does not depend on the size of the workload. 
Let $\overhead$ denote the overhead (which will be a time delay to start). We use $\groupSize$ to denote `group size', which is a parameter to be discussed later in the context of our proposed algorithm. In the sequel we assume a `master' node is employed for centralized coordination of tasks and workers (e.g., the App Master in MapReduce). 
Next, we motivate our approach with an example.

\subsection{A Motivating Example}

\begin{figure}
	\reducebelowcaptionskip
	\centering\includegraphics[width=1\columnwidth]{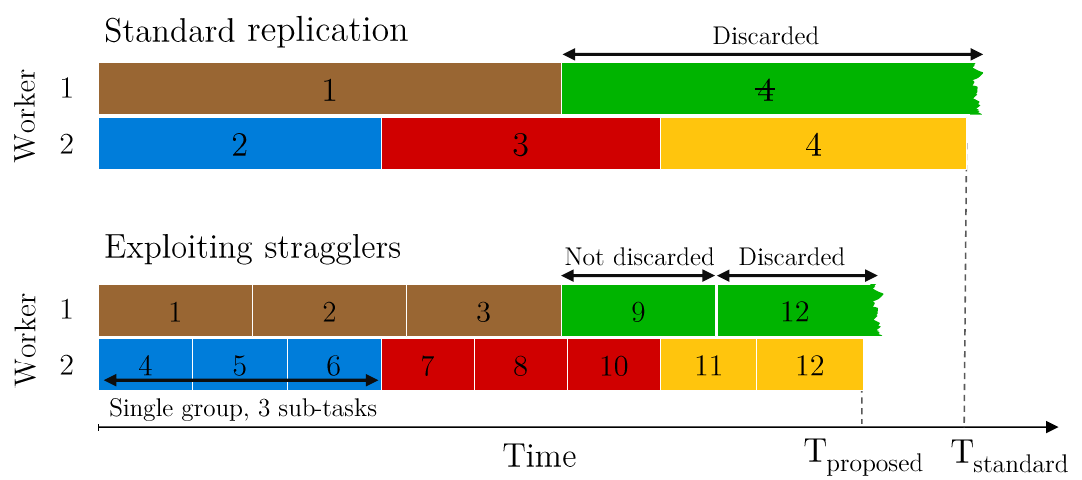}
	\caption{An example that motivates the proposed algorithm.}
	\label{fig:illustration}
\end{figure}

Consider standard replication with two workers and 4 equal size tasks. Assume that worker 2 is faster than worker 1, i.e., worker 1 is a straggler. \fig{illustration} (top part) presents how tasks may be assigned by standard replication. Blocks of different colors indicate tasks. Worker 2 processes tasks faster than worker 1, and the per task processing time for a worker remains constant (e.g., the processing times for tasks 2, 3 and 4 on worker 2 are equal). 
Worker 1 is the first to start processing task 4, and task 4 is replicated on worker 2 as well. 
An important observation in \fig{illustration} is that even though worker 1 completes a significant portion of task 4, all that effort is discarded since worker 2 is the first to finish task 4. (Discarded tasks are indicated with a jagged right-side edge and a strike-through label.) Instead of discarding all the work done by worker 1 on task 4, we would like to be able to utilize some portion of it. 

Now, consider the second algorithm illustrated in \fig{illustration} (bottom part). In this algorithm a worker processes a group of smaller tasks (3 tasks per group in the example). Tasks that share a color belong to the same group (e.g., tasks 4, 5 and 6, and, tasks 10, 11 and 12). To facilitate comparison, we keep the size of the overall workload identical to the top part. For example, note that the width of task 1 in top part is equivalent to the sum of widths of tasks 1, 2, and 3 in the bottom part (the same is true for other tasks, e.g., task 2 in top part is equivalent to the sum of widths of tasks 4, 5, and 6 in the bottom part). Thus, the bottom part allows more granularity of tasks, while keeping the sizes of the workloads identical. Notice that worker 1 stops processing task 12 (and discards work) once worker 2 finishes processing task 12. The important observation is that the work done by worker 1 on task 9 will not get discarded. This contrasts with the discarding of all the work done by straggling worker 1 in standard replication (i.e., task 4 in green color). 

This example demonstrates that by increasing the granularity of work assigned to workers it is possible to exploit work performed by stragglers. 
However, in practice, the task size is not always a parameter that can easily be changed (e.g., the task may be applying a function to an image, in which case the size of the task is not in user's control). The second algorithm, as visualized in the bottom part in \fig{illustration}, provides an alternative. Instead of \emph{splitting} work assigned, the second algorithm achieves a a similar effect by \emph{grouping} multiple tasks. The granularity of workloads assigned to workers increase, as groups consist of multiple tasks. This is the motivation of the group replication approach. 
We demonstrate through numerical experiments presented in Sections~\ref{secn:simulateexp} and~\ref{secn:ec2exp} (and as also evident in \fig{yarntestresults}) that there exists a range of group sizes that yield reduced processing times.

\subsection{Formulating the Proposed Method}

Our approach is to group tasks when assigning them to workers \emph{and} to increase the frequency of worker updates. We note that these modifications do not require any new inputs from the user such as changes to the user-defined task processing logic. To this end we envisage a generalization of standard replication that we summarize below. 

\begin{enumerate}[label=(\alph*)]
\item\emph{Grouping of tasks}: 
The master concatenates multiple tasks into a group and assigns them to a worker. We refer to this as a `task group' or simply a `group'. The number of tasks in a group is a hyper-parameter in the algorithm, denoted by $\groupSize$.

\item\emph{Notification of completed tasks}:
Workers processes tasks within groups sequentially. Workers notify the master immediately after completing each task within a group. 

\item\emph{Skipping completed tasks}:
The master can instruct a worker to skip over a task in the group. Upon receiving a notification of the completion of a task from a worker, the master signals other workers that are holding groups which contain the same task to skip over that task. 
The master can so instruct the worker either before the worker starts processing that particular task or in middle of processing of the task. 
\end{enumerate}

The task group described in point (a) may consist of both unreplicated and replicated tasks. The worker starts processing replicated tasks only after processing all unreplicated tasks. To determine which tasks must be replicated, one can make use of one of the existing speculation methods discussed in \secn{relatedworkmr}.

\subsection{Proposed Algo. vs. Replication-with-grouping-only} \label{secn:mapredgroupingonly}

The time savings we obtain with the proposed algorithm can be attributed to two sources. The first is reduction in overhead. The second is exploitation of stragglers. Grouping is a commonly used approach to reduce worker initialization overhead. 
The group functionality described in point (a) above already exists in certain implementations of MapReduce. For example, in Hadoop the user may employ the `Combine File Split' feature that creates a sub-collection of input files. Such grouping reduces the overhead $\overhead$ incurred when assigning tasks to a worker, since the overhead incurs only once per group. 
To quantify how much of the time savings of the proposed algorithm is due to reduction of overhead, we implement `replication-with-grouping-only', a version of the algorithm that implements task grouping without increasing the frequency of worker updates. This algorithm implements functionalities described in points (a) and (c) that describe the grouping aspect. With regard to point (b), while the workers sequentially processes tasks within groups, a worker notifies the master only after completing \emph{all} tasks within the group, i.e., the worker does not notify the master after completing each task. This implementation excludes the increased worker communication, hence the name replication-with-grouping-only. Not that when $\groupSize=1$ replication-with-grouping-only reduces to standard replication. 

Grouping functionality does not alone address the straggler issue. To address stragglers also requires the functionality described in point (b), the increased frequency of worker updates. We verify in \secn{simulateexp} that this combination of grouping and increased communication in the proposed algorithm offers considerable time savings when compared both to standard replication and to replication-with-grouping-only. 
To quantify the fraction of time savings due to overhead reduction, 
in \secn{simulateexp} we conduct specially designed experiments on a simulated cluster. A simulated cluster allows us to vary the overhead $\overhead$ while quantifying the time savings offered by the proposed algorithm. Our experiments demonstrate that the proposed algorithm offers considerable time savings even when the overhead $\overhead$ is negligible.

\subsection{Visual Comparison of Replication Methods}
\begin{figure}
\reducebelowcaptionskip
\centering\includegraphics[width=1\columnwidth]{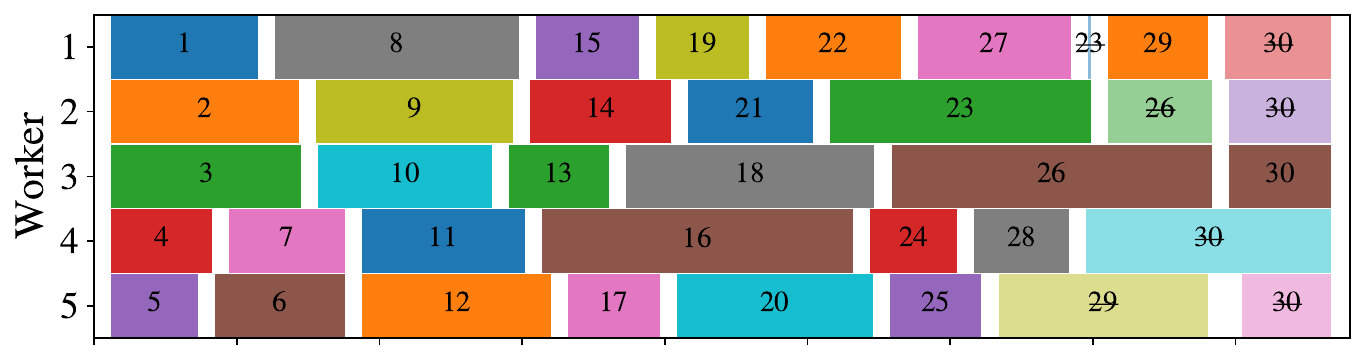}
\\ \vspace{-0.6ex}
\centering\includegraphics[width=1\columnwidth]{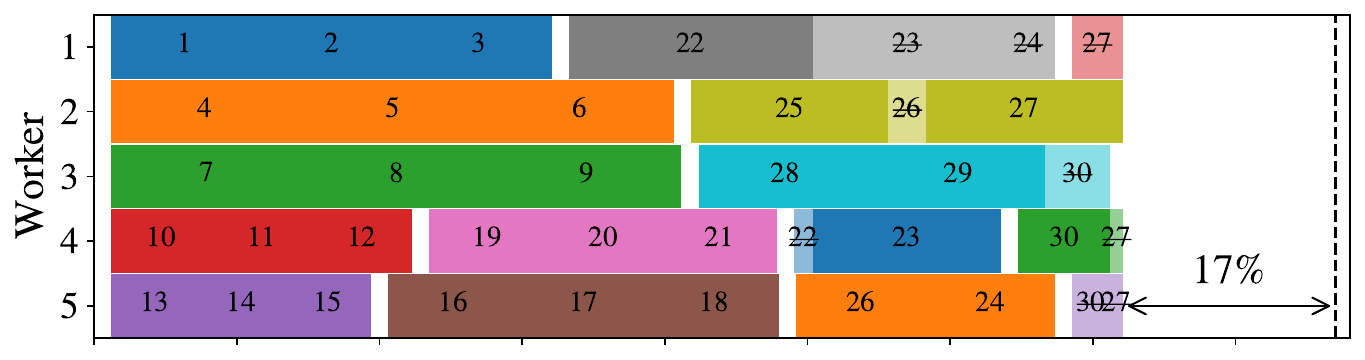}
\\ \vspace{-0.6ex}
\centering\includegraphics[width=1\columnwidth]{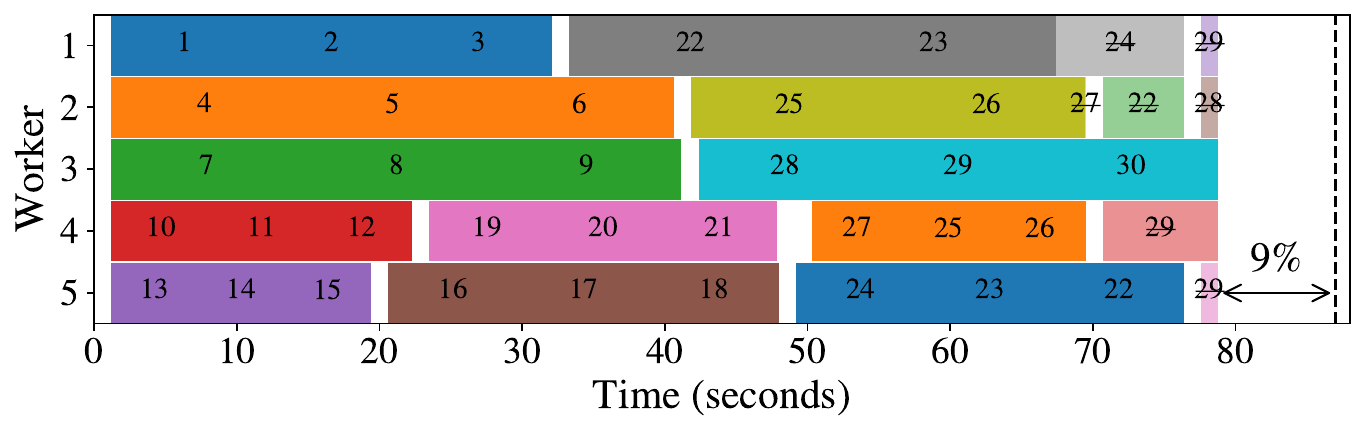}
\caption{Visualizing task assignments in standard replication (top), proposed method (middle), and replication-with-grouping-only (bottom). For the last two, group size $\groupSize=3$. 
}
\label{fig:amrvanillanorep}
\end{figure}

\fig{amrvanillanorep} visualizes various approaches. The goal of \fig{amrvanillanorep} is providing a quick visual comparison of different methods. Therefore, we point the reader to \secn{simsetup} for details on how these results are generated. In all sub-figures $\numWorkers=5$, $\numSplits=30$ and $\overhead=1.2$ seconds. The straggler statistics across the three figures are identical. 

The top sub-figure visualizes standard replication. The white space to the left of each task represents the overhead duration $\overhead$. The processing time of tasks (indicated by the width of a block) vary due to the presence of stragglers. If there were no stragglers in the system and all tasks were of equal size, they would take an equal amount of time. I.e., each worker would process $6$ tasks, and all workers would finish at the same time. 

In the top sub-figure, when replicating a task, we uniformly select a task from the available ones. For example, by the time worker 2 finishes task 23, only 3 tasks remain incomplete, i.e., task 26, 29 and 30. All three of these tasks are being processed by other workers. At this point worker 2 replicates one of the three tasks. In the example depicted worker 2 replicates task 26. Similarly, after finishing task 27, worker 1 replicates task 23, as all other tasks are either completed or in progress (note that worker 4 begins task 30 momentarily before worker 1 begins task 23). The master collects the completed work only from the first worker to finish a task and, once this occurs, all other workers processing the same task are instructed to stop the processing that task. For example, worker 3 is the first to finish task 26, and once worker 3 completes it, worker 2 stops processing task 26 (as indicated by the strike-through label, which shows that the task's processing was prematurely stopped). Observe that, although worker 5 completes a significant portion of task 29, the efforts of worker 5 are discarded once worker 1 completes task 29. The same observation is true about task 30 of worker 4. Discarding work done by some of the workers is a waste of resources. In contrast, with our proposed algorithm we attempt to \emph{exploit} \emph{all} the work done by workers. 

The middle sub-figure depicts the functioning of the proposed method when $\groupSize=3$. Recall that the idea is first to group tasks when assigning tasks to workers, and second to increase the frequency of worker updates. Tasks belonging to the same group have the same color (e.g., tasks 1, 2 and 3 of worker 1). Some groups are seen as consisting of only two or one task, e.g., the group with tasks 22 and 23 of worker 4. This is because, even though the original group assignment consists of 3 tasks, some of the tasks may have already been completed by other workers. Such tasks are not shown in the illustration to avoid clutter. In the same example, worker 4 may have been assigned tasks 22, 23 and 25. Task 25 does not show up since it was completed by worker 2 before worker 4 starts processing task 25. When there exists multiple replicated tasks within the same group, the order of execution is randomized. This is the reason for worker 5 executing task 26 prior to task 24. We provide the full specifications of master's and worker's functioning in the next section. Note that the strategy in standard replication is a special case of the proposed algorithm, when the group size $\groupSize$ is set to $\groupSize=1$ (i.e., each group consists of a single task). 

The bottom sub-figure depicts replication-with-grouping-only when $\groupSize=3$. Both worker 1 and 5 end up completely processing task 22 and 23. Even though grouping reduces worker overhead, the lack of communication between workers leads to resource wastage. In comparison, in the proposed method, worker 1 stops processing task 23 once worker 4 completes task 23. In \secn{simulateexp} we demonstrate that for large $\groupSize$ the algorithm that implements replication-with-grouping-only may increase the processing time compared to standard replication. The key takeaway from \fig{amrvanillanorep} is that the proposed method reduces resource wastage and, in turn, it yields the most time savings compared to the other two methods.

\subsection{Master's and Worker's Functioning}

\newcommand{\setL}[1]{\calL_{\text{#1}}}
\newcommand{\setG}[1]{\calG_{\text{#1}}}
\newcommand{\setsize}[1]{|#1|}

\input{alg1_master}

In \algo{appmaster} and \algo{mapper} we detail the functioning of the master and a worker. In what follows $\setsize{\cdot}$ denotes the cardinality of the argument. In \algo{appmaster}, master maintains two lists. List $\setL{U}$ is a list of \emph{unassigned} tasks to be processed by workers as they become available. List $\setL{A}$ is a list of already \emph{assigned} tasks currently being processed by workers. In lines \ref{algo:line:masterwhilestart} to \ref{algo:line:masterwhileend}, the master keeps checking for worker updates and calls \texttt{AssignGroupToWorker} and \texttt{MarkTaskAsComplete} functions. \texttt{AssignGroupToWorker} assigns task groups to workers and updates $\setL{U}$ and $\setL{A}$ accordingly. When the master receives notification that a task has been completed by some worker, \texttt{MarkTaskAsComplete} instructs other workers to skip over that particular task. 

Next we describe how task replication occurs in \texttt{AssignGroupToWorker}. In lines \ref{algo:line:assignstart} to \ref{algo:line:assignend} tasks in $\setL{U}$ are assigned to workers for the first time. In lines \ref{algo:line:replstart} to \ref{algo:line:replend} tasks in $\setL{A}$ that are already being processed at workers are replicated. When forming a task group the elements in $\setL{U}$ have priority over elements in $\setL{A}$. 
The parameter $R$ in \ref{algo:line:paramR} is the space remaining in the task group after adding elements from $\setL{U}$. Note that task replication happens only if $R>0$. When a worker finishes its current group of tasks, it is assigned $\calG$, the concatenation of $\setG{U}$ (a subset of $\setL{U}$) and $\setG{A}$ (a subset of $\setL{A}$). 

In line \ref{algo:line:uniform} we select $S$, the task to replicate next. In our experiments in \secn{simulateexp} and \secn{ec2exp} we select $S$ as follows. In line \ref{algo:line:setdiffJ} $\calJ$ is the set of tasks that can be replicated but are not yet added to the current task group. We select the least replicated task in $\calJ$ for $S$. If there exist multiple candidates, i.e., tasks that have been replicated the same number of times, we pick $S$ uniformly at random. This reduces the likelihood that different workers work on the same replicated task at the same time. This simple strategy for selecting $S$ provides a baseline to make a fair comparison between standard replication and the proposed algorithm. If need be, one could select $S$ using one of the speculation-based techniques discussed in \secn{relatedworkmr}. We anticipate even greater time savings for both standard replication and the proposed algorithm through the application of such speculation-based techniques. 

\algo{mapper} describes how the worker functions. The input to the worker is the task group $\calG$. In essence, the worker sequentially processes tasks in $\calG$, while skipping over tasks the master has signalled to skip. 
We emphasize the user only needs to define the processing logic of $S$, which is required by standard replication in any case.

\input{alg2_worker}

We note that some operations can be made more efficient. For example, the broadcast operation on line~\ref{algo:line:stopskip} in \algo{appmaster} can be replaced by a multicast operation by maintaining a list of workers $S$ is assigned to. While to keep presentation simple we avoid including such improvements in \algo{appmaster} and \algo{mapper}, we do incorporate such improvements in the implementation of the proposed algorithm discussed in \secn{ec2exp}.

In the following sections, we empirically analyze the performance of the proposed algorithm and compare it to standard replication. We conduct two types of experiments. 
In \secn{simulateexp} we present experiments conducted in a simulated cluster. 
These experiments help us understand how much of the time savings is due to the exploitation of stragglers, and how much is due to the reduction of worker overhead. 
In \secn{ec2exp} we present details of the experiments conducted with Apache Hadoop on Amazon EC2. These experiments complement and verify our findings in \secn{simulateexp}.

\input{numerical_sim}
\input{numerical_ec2}

\section{Conclusions and Future Work} \label{secn:nextstepsmrf}
In this paper, we proposed an algorithm for exploiting work performed by stragglers in distributed computing systems. Our contributions are two-fold. First, we introduced a method to exploit stragglers by increasing the granularity of work assigned to workers. We achieved this first by grouping tasks when assigning them to workers, and second by increasing the frequency of worker updates. These changes allow the proposed algorithm to make use of the progress made by stragglers rather than simply disregarding their work. Task grouping is commonly used to reduce worker initialization overhead. We conducted experiments to tease out what fraction of the time reduction is due to reductions in overhead and how much is due to exploitation of stragglers. Our experiments demonstrated that the time savings are due to both.

As our second contribution, we introduced a computational framework in which we can conduct repeatable experiments of a distributed system that suffers from stragglers. The idea is first, to obtain a set of arbitrarily induced straggler profiles. We accomplished this by artificially loading CPU cores and by measuring the completion time of a set of tasks. Second, given a desired target straggler histogram of a real system, we induced a convex combination of straggler profiles to match the target histogram, i.e., a mixture model of stragglers. This method allows us to study different algorithms and parameter settings in a controlled manner while keeping straggler characteristics are reproducible. 

In the proposed algorithm we introduced a parameter we call the group size. One possible next step in this theme of works is developing an analysis that characterizes the performance improvement in terms of the group size parameter. Such an analysis will prove beneficial to the practitioners of the proposed algorithm in determining the optimal group size for a given set of tasks. We leave this as future work.

%% file: notationSpecific.tex
\pragmaonce  

\newcommand{\numWorkers}{W}
\newcommand{\numSplits}{S}
\newcommand{\overhead}{H}
\newcommand{\expShift}{T}
\newcommand{\expLambda}{\lambda}
\newcommand{\groupSize}{G}

%% file: alg1_master.tex
\begin{algorithm}
	\caption{Master functioning} \label{algo:appmaster}
	\SetAlgoVlined \LinesNumbered \SetKwInOut{Initialize}{initialize}
	\SetKwInOut{Input}{input} 
	\SetKwFunction{FAssign}{AssignGroupToWorker}
	\SetKwFunction{FSplitComplete}{MarkTaskAsComplete}
	
	\Input{$\setL{U}=$ list of tasks, $\groupSize=$ group size \;}
	\Initialize{$\setL{A}=$ empty list \;}
	\While{$\setsize{\setL{U}}>0$~\algorithmicor~$\setsize{\setL{A}}>0$}{ \label{algo:line:masterwhilestart}
		\If{a worker $M$ finishes assigned group}{
			call \FAssign{$M$} \;
		}
		\If{a worker notifies that task $S$ is complete}{
			call \FSplitComplete{$S$} \label{algo:line:masterwhileend} \;
		}
	}
	
	\SetKwProg{Fn}{Function}{:}{}
	\Fn{\FAssign{$M$}}{
		$\setG{U}=$ empty list \label{algo:line:assignstart} \;
		\While{$\setsize{\setL{U}}>0$~\algorithmicand~$\setsize{\setG{U}}<\groupSize$}{
			let $S$ be the first element in $\setL{U}$ \;
			remove $S$ from $\setL{U}$ \;
			append $S$ to both $\setG{U}$ and $\setL{A}$ \label{algo:line:assignend} \;
		}
		$R=G-\setsize{\setG{U}}$\label{algo:line:paramR},
		$\setG{A}=$ empty list \label{algo:line:replstart} \;
		\While{$\setsize{\setL{A}}>0$~\algorithmicand~$\setsize{\setG{A}}<R$}{
			compute the set difference $\calJ=\setL{A}\setminus \setG{A}$ \label{algo:line:setdiffJ} \;
			\If{$\calJ$ is empty}{
				break loop \;
			} \Else {
				pick $S$ from $\calJ$, the task to replicate next \label{algo:line:uniform} \;
				append $S$ to $\setG{A}$ \label{algo:line:replend} \;
			}
		}
		let $\calG$ be the concatenation of $\setG{U}$ and $\setG{A}$ \;
		assign $\calG$ to $M$ to process with \algo{mapper} \;
	}
	
	\Fn{\FSplitComplete{$S$}}{
		remove $S$ from $\setL{A}$ if exits \;
		signal to all workers to stop/skip processing $S$ \label{algo:line:stopskip} \;  
	}
\end{algorithm}

%% file: alg2_worker.tex
\begin{algorithm}
	\caption{Worker functioning} \label{algo:mapper}
	\SetAlgoVlined \LinesNumbered \SetKwInOut{Initialize}{initialize}
	\SetKwInOut{Input}{input} \SetKwInOut{Output}{output}
	\Input{$\calG=$ list of tasks \;}
	\While{$\calG$ is non-empty}{
		remove the first element in $\calG$ and call it $S$ \;
		set $\texttt{IsTaskComplete}$ boolean to $\texttt{False}$ \;
		\While{$\texttt{True}$} {
			apply next step of processing to $S$ \label{algo:line:nextrecord} \;
			\If{master signalled to skip processing $S$}{
				break loop \;
			}
			\If{processing $S$ is complete} {
				set $\texttt{IsTaskComplete}$ to $\texttt{True}$ \;
				break loop \;
			}
		}
		\If{$\texttt{IsTaskComplete}$ is $\texttt{True}$} {
			notify master that processing $S$ is complete \;
		}
		remove from $\calG$ tasks the master has signalled to skip processing \;
	}
\end{algorithm}

%% file: numerical_sim.tex

\section{Simulation-based Experiments} \label{secn:simulateexp}

In this section we conduct 
experiments to measure the time savings effected by the proposed algorithm as a function of varying cluster-dependent parameters. 
In the following sections we present results obtained by varying the overhead $\overhead$, the straggler variability, and the number of tasks in the workload. 
Simulation-based experiments are important as some parameters (e.g., the overhead $\overhead$) are not easy to change under the effect of real-world clusters.

\subsection{Simulation Setup} \label{secn:simsetup}
To maintain simplicity in the simulation setup, we assume the tasks are of uniform size. (We will explore tasks of varying sizes in experiments conducted on Amazon EC2, presented later in \secn{ec2exp}.) Let $X$ be a random variable that denotes the time required for processing a single task. To keep the simulation model simple, we make two assumptions about the processing speeds of workers. First, we assume that the processing times of tasks within a group are equal, i.e., each task within a group takes an equal amount of processing time. While in practice the processing speed of a worker is independent of the group size, this assumption provides us with a way to easily model temporal consistency in straggler behaviour over short periods of time. Second, we assume that $X$ is sampled independently across groups. This assumption enables us to model the randomness of processing speeds workers over long periods of time. 

In the simulation-based experiments we simulate different straggler statistics by sampling $X$ from PDFs we specify next. 
Our probability distribution is the `shifted exponential'. 
For positive scalars $\lambda$ and $\expShift$, let the probability density function (PDF) of a shifted exponential random variable be $f_{\lambda,\expShift}(x)$, where 
$f_{\lambda,\expShift}(x) = \lambda\exp{\left(-\lambda\left(x-\expShift\right)\right)}$ if $x\geq\expShift$ and 
$f_{\lambda,\expShift}(x) = 0$ otherwise.
The results in \fig{amrvanillanorep} are generated by sampling $X$ from $f_{\lambda,\expShift}$ with $\lambda=0.15$, $\expShift=6$ seconds. For this $\overhead=1.2$ seconds and $\numSplits=30$, and we recall that the top and bottom sub-figures in \fig{amrvanillanorep} respectively correspond to $G=1$ and $G=3$. 
In $f_{\lambda,\expShift}$, $\expShift$ is the minimum amount of time required to complete a task. 
The $\lambda$ parameter controls the additional processing time. A smaller $\lambda$ leads to higher variance and higher straggler variability. 

\begin{figure}
	\reducebelowcaptionskip
	\centering\includegraphics[width=1\columnwidth]{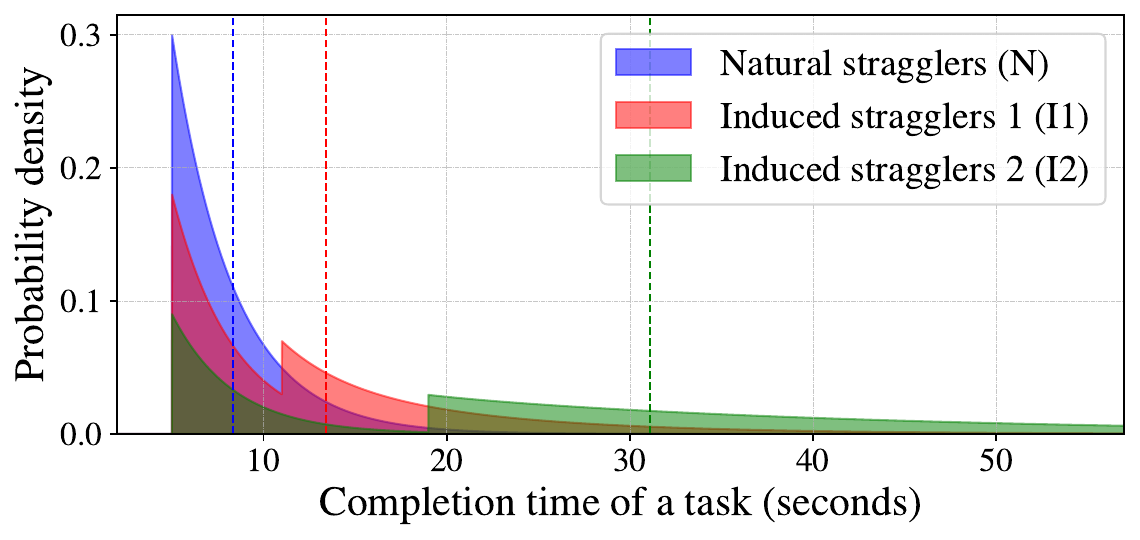}
	\caption{The PDFs for $X$, the time required to process a single task. For each PDF the dashed vertical line indicates the expected completion time.}
	\label{fig:simstragglerpdf}
\end{figure}

For simulating different straggler statistics, we use a mixture of shifted exponential distributions by taking convex combinations of $f_{\lambda,\expShift}$ for different combinations of $\lambda$ and $\expShift$. 
Considering \fig{simstragglerpdf}, the PDFs (N), (I1) and (I2) correspond to 
$f_{0.3,5}(x)$,
$0.6f_{0.3,5}(x)+0.4f_{0.1,11}(x)$,
and
$0.3f_{0.3,5}(x)+0.7f_{0.04,19}(x)$. 
Each mixture of $f_{\lambda,\expShift}$ means, e.g., in (I1), that we sample from $f_{0.3,5}$ with probability $0.6$, and from $f_{0.1,11}$ with probability $0.4$. 
We obtained the PDFs in \fig{simstragglerpdf} by tuning the combinations of parameters to match empirical straggler statistics we observe in Amazon EC2 experiments (cf.,\secn{ec2exp}). 
Respectively, we tune the PDFs (N), (I1) and (I2) to match their expected values (dashed vertical lines in \fig{simstragglerpdf}) with the averages of the empirical task completion time obtained with natural and induced stragglers in our EC2 experiments (cf. dashed vertical lines in \fig{completiontimehist}). 
Note that the variance of PDFs increases in order (N), (I1) and (I2). The results presented in \fig{amrtimevsoverhead} through \ref{fig:amrmontecarlolambda} are generated with this simulation setup. In all figures, each plot is averaged over 10,000 repetitions; thus, the variance of a plots is negligible. 

\subsection{Time Savings vs. Overhead}  \label{secn:amrtimevsoverhead}
\begin{figure*}
	\centering
	\begin{subfigure}{0.339\textwidth}\centering
		\includegraphics[width=1\textwidth]{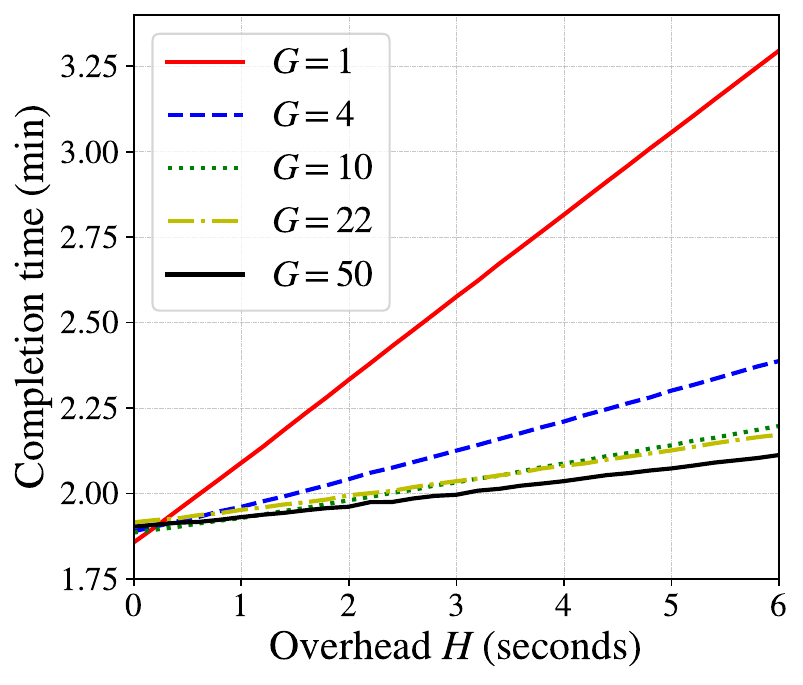}
		\caption{With PDF (N).}
		\label{fig:amrtimevsoverheada}
	\end{subfigure}\hfill%
	\begin{subfigure}{0.32\textwidth}\centering
		\includegraphics[width=1\textwidth]{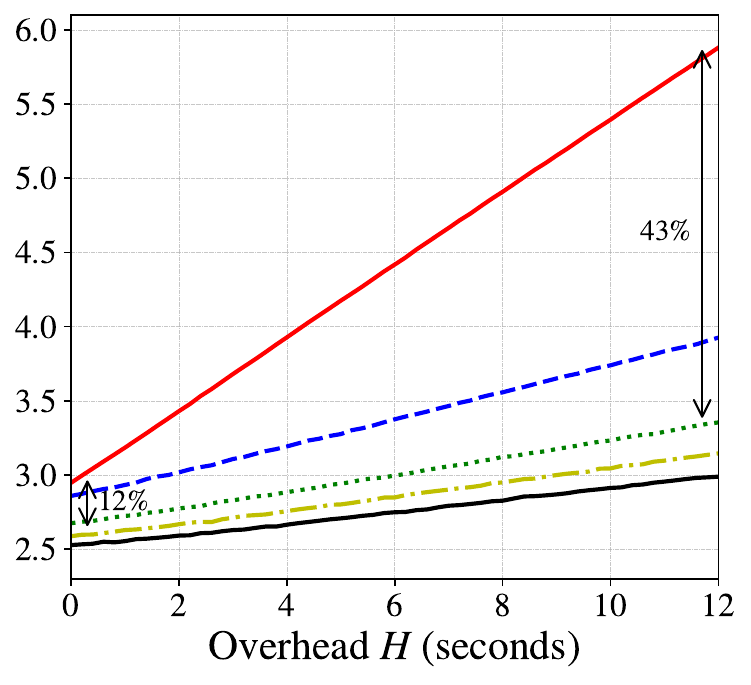}
		\caption{With PDF (I1).}
		\label{fig:amrtimevsoverheadb}
	\end{subfigure}\hfill%
	\begin{subfigure}{0.31\textwidth}\centering
		\includegraphics[width=1\textwidth]{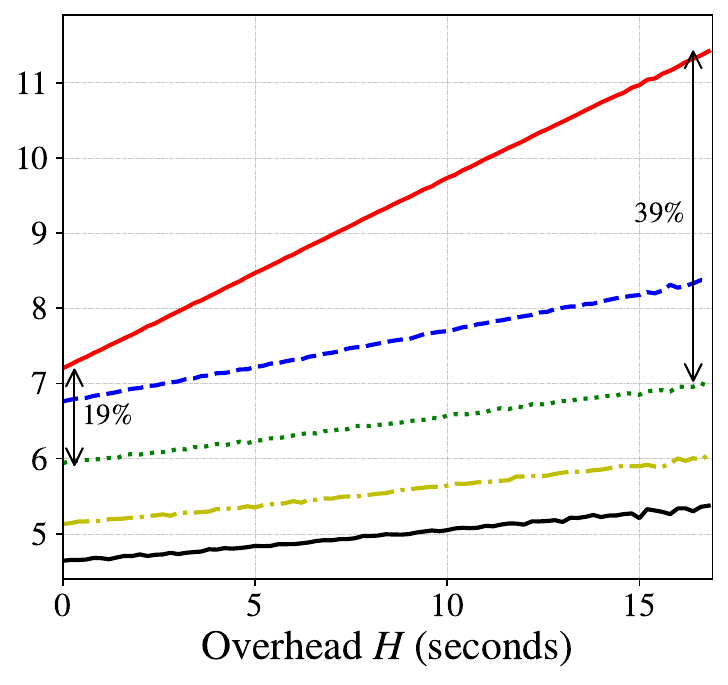}
		\caption{With PDF (I2).}
		\label{fig:amrtimevsoverheadc}
	\end{subfigure}%
	\caption{Completion time vs. overhead with the three straggler PDFs in \fig{simstragglerpdf}.
	In all sub-figures, $\groupSize=1$ plots correspond to standard replication, and the rest correspond to the proposed algorithm. 
	Each plot is generated by varying $\overhead$ from zero to a maximum value. The maximum $\overhead$ is selected to match the completion time with Amazon EC2 results. For example, with (I1) plot in \fig{yarntestresults}, at $\groupSize=1$ it takes around $5.9$ minutes. In \fig{amrtimevsoverheadb} this amount of time is attained at around $\overhead=12$. 
	}
	\label{fig:amrtimevsoverhead}
\end{figure*}

\begin{figure*}
\minipage{0.32\textwidth}
\centering\includegraphics[width=1\textwidth]{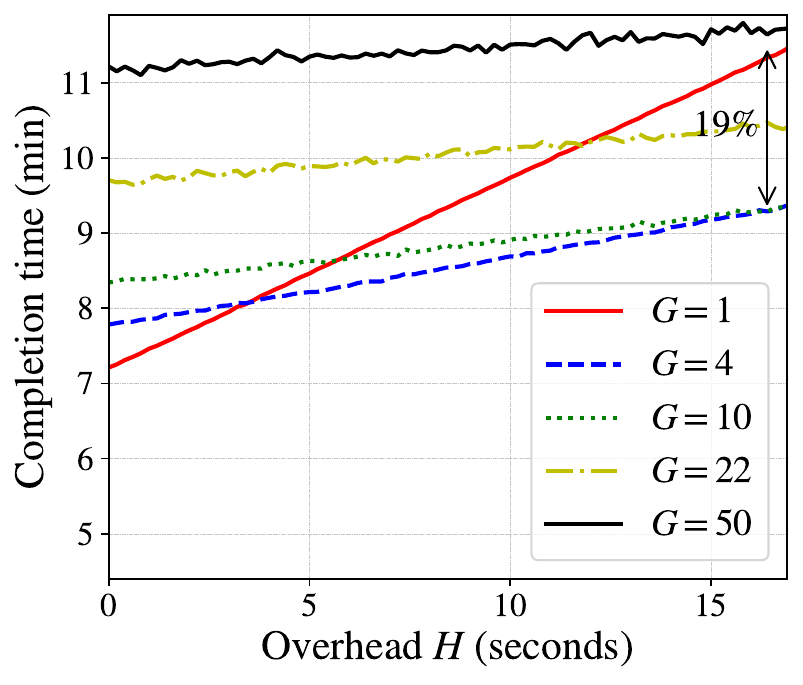}
\caption{\fig{amrtimevsoverheadc} equivalent for replication-with-grouping-only with PDF (I2).}
\label{fig:time_vs_overhead__induced2_grouping_only}
\endminipage\hfill
\minipage{0.32\textwidth}
\centering\includegraphics[width=1\textwidth]{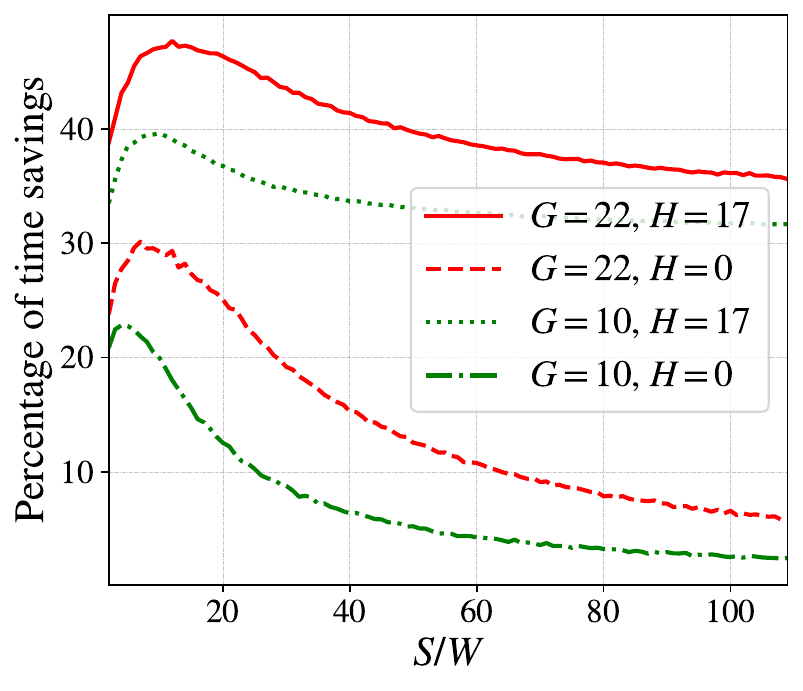}
\caption{Time savings vs. number of tasks to number of workers ratio.}
\label{fig:ratio_vs_savings__induced2}
\endminipage\hfill
\minipage{0.32\textwidth}%
\centering\includegraphics[width=1\columnwidth]{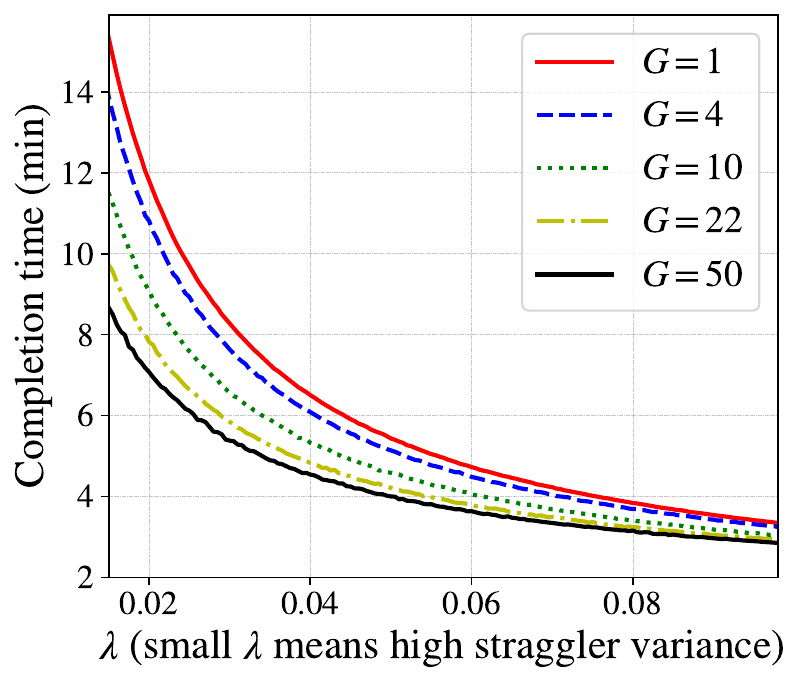}
\caption{Completion time vs. straggler variance parameter for $\overhead=0$. 
}
\label{fig:amrmontecarlolambda}
\endminipage
\end{figure*}

To quantify the time savings due to straggler exploitation, we conduct experiments with each of the PDFs in \fig{simstragglerpdf} and vary $\overhead$. To match the task parameters of the simulations with the EC2-based experiments we present in \secn{ec2exp}, we set $\numSplits=392$ and $\numWorkers=31$. \fig{amrtimevsoverhead} and \fig{time_vs_overhead__induced2_grouping_only} present the results respectively for the proposed algorithm and for replication-with-grouping-only. 
For ease of presentation we discuss \fig{time_vs_overhead__induced2_grouping_only} first and then proceed to \fig{amrtimevsoverhead}.

\fig{time_vs_overhead__induced2_grouping_only} presents the results for replication-with-grouping-only when processing times are distributed according to PDF (I2). Recall that, as discussed in \secn{mapredgroupingonly}, replication-with-grouping-only is equivalent to standard replication when $\groupSize=1$. In \fig{time_vs_overhead__induced2_grouping_only}, some values of $\groupSize$ yield significant time savings when $\overhead$ is large. For example, when $\overhead=17$ both $\groupSize=4$ and $\groupSize=10$ yield time savings of around $19\%$. These savings can be attributed to overhead reduction due to grouping. However, as $\groupSize$ increases, the time savings quickly vanish. For $\groupSize=50$ the completion time exceeds that for $\groupSize=1$. This demonstrates that for large $\groupSize$ the lack of communication between workers makes grouping more hurtful than helpful.

\fig{amrtimevsoverhead} presents results for our proposed algorithm. The three sub-figures correspond to the three PDFs of completion times (N), (I1) and (I2). In all sub-figures the $\groupSize=1$ plots correspond to standard replication. The other plots correspond to our algorithm. 
The three sub-figures in \fig{amrtimevsoverhead} demonstrate that, as straggling becomes more acute (left to right), 
the time savings due to grouping becomes more significant. 
Consider the $\groupSize=10$ plots across the three sub-figures at overhead $\overhead=0$. 
In \fig{amrtimevsoverheada} when straggler variance is low, completion time for $\groupSize=10$ is basically the same (actually slightly increased) as for $\groupSize=1$. 
However, as straggler variance increases, the percentage reduction in completion time for $\groupSize=10$ increases consistently ($12\%$ in \fig{amrtimevsoverheadb} and then $19\%$ in \fig{amrtimevsoverheadc}). 
When $\overhead=0$, any time reduction must be solely due to exploitation of stragglers rather than reduction in overhead (since overhead is zero). \fig{amrtimevsoverhead} therefore demonstrates that the leveraging of both grouping and increased worker communication in the proposed algorithm is a good strategy for exploiting stragglers.

\fig{amrtimevsoverheadc} and \fig{time_vs_overhead__induced2_grouping_only} are directly comparable since both are generated with PDF (I2). 
Since both the proposed algorithm and replication-with-grouping-only reduce to standard replication when $\groupSize=1$, the $\groupSize=1$ plots in both \fig{amrtimevsoverheadc} and \fig{time_vs_overhead__induced2_grouping_only} are identical. For all values of $\groupSize$, the proposed algorithm offers significantly more time savings than replication-with-grouping-only. For example, when $\overhead=17$ we observe $\groupSize=10$ in \fig{amrtimevsoverheadc} offers close to $39\%$ time savings. This is roughly twice the percentage savings offered by replication-with-grouping-only in \fig{time_vs_overhead__induced2_grouping_only}. A similar observation is true for other values of $\groupSize$. Since the proposed algorithm offers considerably more time savings than replication-with-grouping-only, the key takeaway from these experiments is that the proposed algorithm's time savings is due not only to overhead reduction but also to making use of the partial work done by stragglers.

Finally, we conduct an experiment to evaluate how the performance improvement reported in \fig{amrtimevsoverhead} vary as a function of $\frac{\numSplits}{\numWorkers}$, the ratio of the number of tasks to the number of workers. We fix $\numWorkers=31$ and vary the number of tasks in the workload. \fig{ratio_vs_savings__induced2} presents the results. We note that the plots in \fig{ratio_vs_savings__induced2} remain unaffected by changes in the value of $\numWorkers$ as the $x$-axis only represents a ratio. The time to complete a task is sampled from PDF (I2) (cf. \fig{simstragglerpdf}). The plots in \fig{ratio_vs_savings__induced2} show how the percentage of time savings vary for no overhead ($\overhead=0$) and for large overhead ($\overhead=17$), for two values of $\groupSize$, 10 and 22. 
The plots show the percentage of time savings compared to group size $\groupSize=1$ (each for the same overhead). 
In all plots we observe that there exist an optimal $\frac{\numSplits}{\numWorkers}$ for which the time savings of the proposed algorithm are the greatest. The savings diminish in the farthest-right region, i.e., for large $\frac{\numSplits}{\numWorkers}$. This is because task replications happen only at the end of the processing of tasks. Thus any gains due to straggler mitigation materialize only towards the end of processing (the last wave of tasks). As $\frac{\numSplits}{\numWorkers}$ increases, the fraction of time spent processing the last of the tasks decreases. Hence the percentage of time savings due to exploiting stragglers also decreases as $\frac{\numSplits}{\numWorkers}$ becomes large.

\subsection{Time Savings vs. Straggler Variance}
In this section we conduct experiments by setting $\overhead=0$ and by varying the straggler variance. We use the PDF (N) (cf. \fig{simstragglerpdf}) and change the $\lambda$ parameter. In other words we sample task completion times from $f_{\lambda,5}(x)$ for varying $\lambda$. The remaining parameters are the same as in \secn{amrtimevsoverhead} (i.e., $\numSplits=392$ and $\numWorkers=31$). \fig{amrmontecarlolambda} presents the results. We observe that, as straggler variance increases, large $\groupSize$ yields more savings compared to $\groupSize=1$. In the small $\lambda$ regime, increasing $\groupSize$ helps reduce the completion time significantly. Since $\overhead=0$ in these experiments, the results in \fig{amrmontecarlolambda} reaffirm our conclusion in \secn{amrtimevsoverhead} that the significant improvements from the proposed algorithm result from the exploitation of stragglers.

One may note that in \fig{amrmontecarlolambda} increasing $\groupSize$ always decreases the completion time (an effect also observed in \fig{amrtimevsoverhead}). This is contrary to the observation in \fig{yarntestresults}, where the completion time plateaus (or increase slightly) as $\groupSize$ increases. We conjecture that this dissimilarity is due to the differences between the simulated and real world clusters. There will be unmodeled aspects of real clusters not reflected in simulation. Examples include delays due to network congestion and variability communication speeds due to topology of workers. Even though we do not model these types of effects of real clusters in our simulated experiments, the two settings produce comparable results for small enough $\groupSize$ (e.g., $\groupSize$ up to around $10$). This range of $\groupSize$ is the range that is most relevant to the conclusions we draw in \secn{simulateexp}, since in our EC2 experiments (cf. \fig{yarntestresults}) the completion time starts to plateau for $\groupSize$ larger than around $10$.

\subsection{Elastic Computing Clusters}
\begin{figure}
\reducebelowcaptionskip
\centering\includegraphics[width=1\columnwidth]{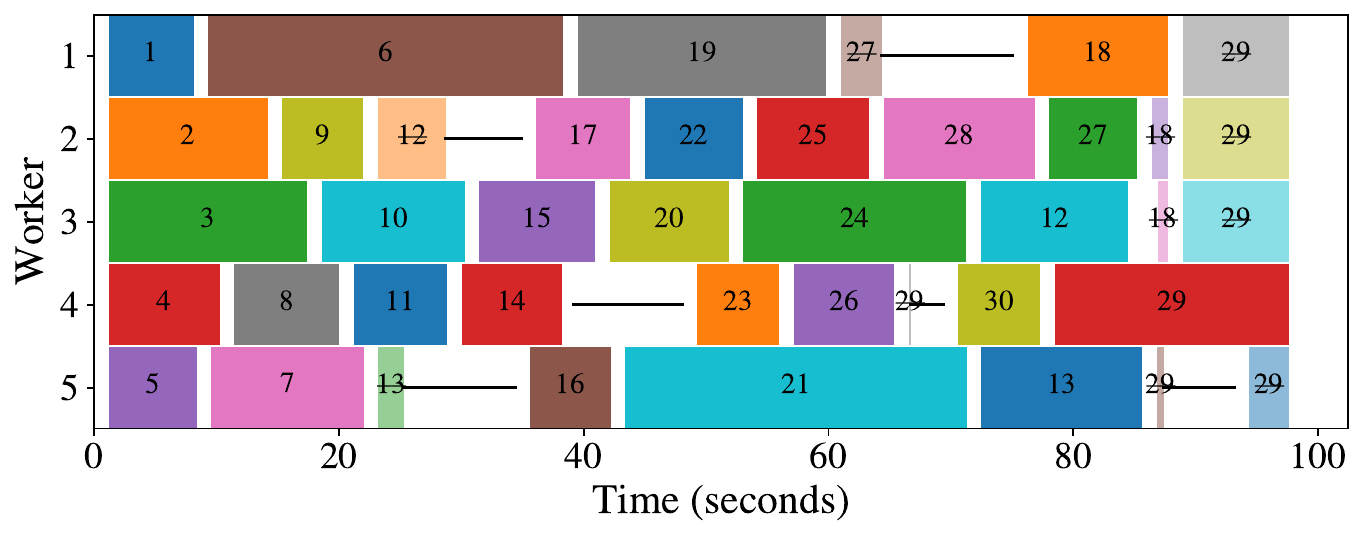}
\caption{Visualizing task assignments when workers are allowed to leave and join. The simulation parameters are similar to the ones used in generating \fig{amrvanillanorep}. The thick black horizontal lines indicate the periods where workers are unavailable.} \label{fig:workers_leave_join}
\end{figure}

We also conduct experiments for a simulated `elastic' compute cluster. In this type of clusters `non-dedicated' workers are present. Non-dedicated workers are allowed to leave or join the cluster at short or no notice. Services such as Amazon EC2 Spot and Microsoft Azure Batch are examples of cloud computing services that provide such non-dedicated workers. \fig{workers_leave_join} visualizes how the task assignments may appear with non-dedicated workers. The thick black horizontal lines indicate the periods in which workers are unavailable. 
We determine the leaving and joining times of workers as follows. Let $T_{\text{A}}$ and $T_{\text{U}}$ be exponential random variables with parameters $\lambda_{\text{A}}$ and $\lambda_{\text{U}}$ respectively. Here, $T_{\text{A}}$ is the duration a worker stays available in the cluster before leaving (measured as the time since it last joined the cluster). Similarly, $T_{\text{U}}$ is the duration a worker stays unavailable before re-joining (measured as the time since it left the cluster). We set $\lambda_{\text{A}}=0.01$ and $\lambda_{\text{U}}=0.1$, and sample $T_{\text{A}}$ and $T_{\text{U}}$ independently across workers. \fig{workers_leave_join} is generated in this manner.

We repeated the experiments presented in \secn{amrtimevsoverhead} with non-dedicated workers and generated results equivalent to those presented in \fig{amrtimevsoverhead}. The results we generated are broadly similar to those in \fig{amrtimevsoverhead}. The completion times increases due to the unavailability of workers. However, the order of the plots and the conclusions remained the same as those made in \secn{amrtimevsoverhead}. Hence, We do not present the \fig{amrtimevsoverhead}-equivalent plots. Our experiments demonstrate that the proposed method is also useful for use in elastic compute clusters. 

%% file: numerical_ec2.tex

\section{Amazon EC2-based Experiments} \label{secn:ec2exp}

In this section we discuss our Amazon EC2-based experiments. These experiments complement and verify our findings in \secn{simulateexp}. We employ the Apache Hadoop software library to implement the proposed algorithm. Since our implementation is quite similar to how the MapReduce algorithm operates in Hadoop, we begin with a brief introduction to the MapReduce framework using the `word count' problem. 

\subsection{MapReduce Framework} \label{secn:mrexplain}
The `word count' problem starts with a large input file containing textual data, and the goal is to prepare a dictionary of word counts (frequencies) of the words in the file. In practice, such large files are stored in a distributed manner. The file is partitioned into many small parts, and each part is stored in multiple compute nodes. The MapReduce framework operates in two phases: the map phase and the reduce phase. In each phase the user provides two problem-specific inputs. The functioning of the two phases and the role of the user's inputs can be understood as follows. 

In the map phase the master prepares a single queue of tasks (`splits' in MapReduce terminology) to be processed. Each worker, referred to as a mapper in this phase, removes the next task in the queue and processes it. This process continues until the queue is empty. The input required from the user is a function that produces the word counts of a text file. Processing a task simply amounts to applying the user-defined function to the file. Observe that a text file gets mapped to a dictionary of word counts, hence the name `map phase'. The user-defined function is referred to as the `map function'. 

In the reduce phase the workers merge all the dictionaries of word counts to produce one single dictionary. They do this via a user-defined `reduce function'. Similar to mapping, reducing is also performed in parallel. Workers recursively merge two dictionaries at a time until a single dictionary of word counts is obtained. Reducing multiple outputs to a single output prompts the name `reduce phase'. Workers in this phase are also referred to as `reducers'. 

\subsection{MapReduce Implementation}
The task processing model we discuss in \fig{task_stack} is quite similar to how the map phase operates in MapReduce. There exists a version of MapReduce implemented in Hadoop, in which only standard replication is implemented. Since the proposed algorithm generalizes standard replication, one may be able to modify the existing MapReduce implementation in Hadoop to implement our proposed algorithm. However, we found it challenging to modify the Hadoop MapReduce implementation due to its various optimizations and legacy constraints. Therefore, we implemented a version of MapReduce from scratch on the Hadoop framework. 

In our implementation, we use Hadoop YARN as the resource management framework and HDFS as the distributed file system. 
We implement two main components, `App Master' and `Worker'. The App Master component is responsible for managing task allocation, communication and replication on workers, as outlined in \algo{appmaster}. The Worker component is responsible for executing process tasks as outlined in \algo{mapper}. These two components emulate the corresponding components in the existing MapReduce implementation in Hadoop. In addition, they support use of the group size parameter. 

\subsection{Dataset, Workload and Cluster Details} \label{secn:datasetsec2}
In our experiments we use two publicly available datasets: the Wikipedia dataset\footnote{To download, follow the instructions in the \texttt{Dataset1-50GB} section at \href{https://engineering.purdue.edu/~puma/datasets.htm}{https://engineering.purdue.edu/$\sim$puma/datasets.htm}.} which comprises of textual data, and the LibriSpeech corpus\footnote{Download the \texttt{original-mp3.tar.gz (87G)} file available at \href{https://www.openslr.org/12}{https://www.openslr.org/12}.} which consists of audio recordings in MP3 format. We preprocess the datasets in the following manner to derive three dataset variations. 

\begin{figure}
\reducebelowcaptionskip
\centering\includegraphics[width=1\columnwidth]{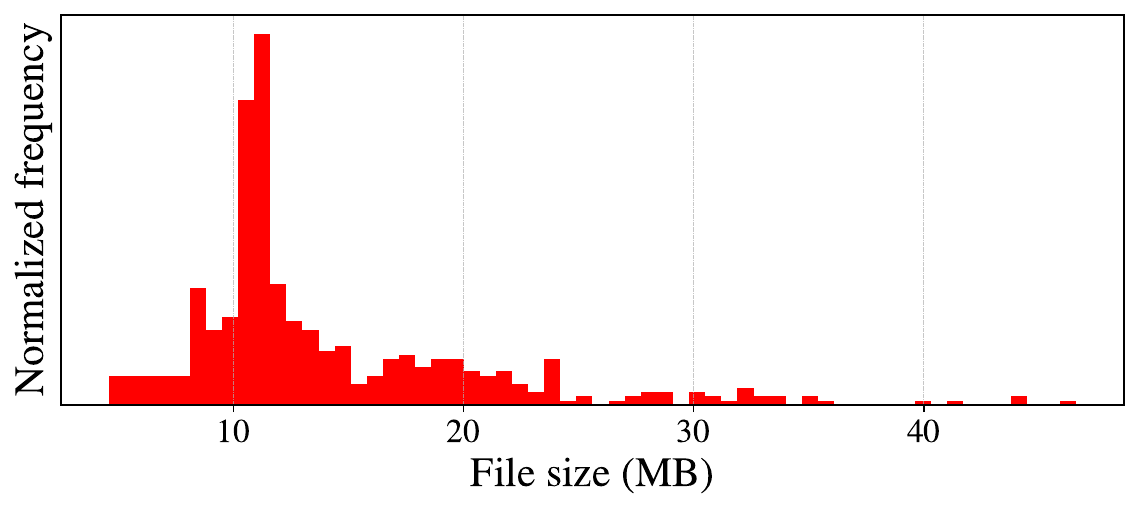}
\caption{File sizes histogram of LibriSpeech-500 dataset.}
\label{fig:yarntestresultslibrispeech}
\end{figure}

\begin{itemize}
\item Wikipedia-Uniform: We preprocess the original Wikipedia dataset by removing files that are smaller than 128~MB and partitioning the remaining files to ensure each does not exceed 128~MB. After preprocessing, the dataset consists of 392 files, each with a uniform size of 128~MB, totaling 50~GB.
\item Wikipedia-Varying: To obtain a dataset with varying file sizes, we further split the files in Wikipedia-Uniform dataset to obtain a total of 2,472 files. The composition of the file sizes in this dataset is \{8 MB: 1280, 16 MB: 640, 32 MB: 320, 64 MB: 160, 128 MB: 72\}, where each entry follows the format `file size: number of files'.  
\item LibriSpeech-500: The original LibriSpeech dataset consists of 5832 MP3 files totalling 85~GB. Due to processing constraints, we use only a subset comprising the first 500 files from the dataset, totalling 6.7~GB. \fig{yarntestresultslibrispeech} presents the histogram of sizes of those 500 files. 
\end{itemize}
We conduct three separate sets of experiments using the three datasets. We use the Wikipedia-Uniform and Wikipedia-Varying datasets for running word count tasks as described in \secn{mrexplain}. We use the LibriSpeech-500 dataset for running audio bit-rate conversion tasks. One such task involves three steps: copying the MP3 file from HDFS to local storage, executing a bit-rate conversion command\footnote{For audio conversion, we use the \texttt{ffmpeg} package with the shell command \texttt{ffmpeg -threads 1 -i <input> -threads:v 1 -map 0:a:0 -b:a 24k <output>}. Consistent with \algo{mapper}, we terminate the command in the middle of its execution if a worker receives the notification that the same task has already been completed by another worker.}, and then copying the output file back to HDFS.

We use a cluster consisting of four Amazon EC2 m5.2xlarge nodes. Each node is equipped with 8~cores, 32~GB memory and 100~GB EBS storage. Workers run in YARN containers each with one core and 4~GB memory. The data is stored in Hadoop's distributed storage, HDFS. In an experiment, one core is allocated to the YARN Application Master. This way, for example, the maximum number of workers that can run simultaneously in the cluster of 4 nodes is $31$, i.e., $\numWorkers=31$.

\subsection{Method of Inducing Random Stragglers} \label{secn:stragglerdetails}
In addition to the experiments involving natural stragglers, we conduct experiments using `artificially induced' stragglers. As the name suggests, we deliberately and randomly slow down some compute nodes while running the proposed algorithm. This helps us mimic a straggler-rich operating environment. 
We induce stragglers by loading the CPU cores using the \texttt{stress-ng} package. The \texttt{stress-ng} command takes two arguments, \texttt{cpu} and \texttt{load}. The first argument controls the number of instances of the stressing program. The second argument controls the percentage load of each program instance. When the \texttt{cpu} value is lower than the number of cores available on a worker node, although not guaranteed, the operating system generally runs the stressing programs on different cores in a non-overlapping manner. To keep the loading on different cores roughly equalized, we set the \texttt{cpu} value to a multiple of the number of cores available in the nodes. We conducted the following experiments to quantify the effect of \texttt{stress-ng} on processing time. 

\subsubsection{Generating Stragglers}
We choose two values for \texttt{cpu} and \texttt{load}, and run \texttt{stress-ng} command on all nodes in the cluster with the chosen two values (i.e., all nodes load CPU cores in a similar manner). At the same time we run the proposed algorithm and measure the completion times of tasks. For example, \fig{plot_hist_grid} presents histograms of task completion times obtained for different values of \texttt{cpu} and \texttt{load} while running word count tasks on the Wikipedia-Uniform dataset. Since we measure only the completion time of a task (and not the completion time of the overall job), the value of group size parameter is not important. In \fig{plot_hist_grid}, the left-most figure is a histogram of computing times obtained without running the \texttt{stress-ng} command (i.e., without inducing stragglers). The following figures correspond to \texttt{stress-ng} command with the \texttt{cpu} value set to $8$, $24$ and $48$. For each of these \texttt{cpu} values we run \texttt{stress-ng} with \texttt{load} value $50$ and $100$. The dashed lines indicate the average completion time. 

The results in \fig{plot_hist_grid} can be summarized as follows. For a fixed \texttt{cpu} value 
a higher \texttt{load} setting shifts the histogram to right. For a fixed \texttt{load} (histograms of the same color across sub-figures), again, a higher \texttt{cpu} setting shifts the histogram to right. In both cases the variance of completion time also changes. The change in variance results from the operating system scheduler. For example, on an 8 core node, for $\texttt{cpu}=24$ we expect three instances of the stressing program to run on one core. However, it is possible that the operating system does not schedule the $24$ stressing program instances uniformly across cores. A larger \texttt{cpu} setting results in a larger number of possible ways to assign the stress programs to cores, which in turn yields a range of possibilities for completion time. This increases the variance of completion time on different cores.

\subsubsection{Combining Multiple Straggler Profiles}
\begin{figure*}
\reducebelowcaptionskip
\centering\includegraphics[width=1\textwidth]{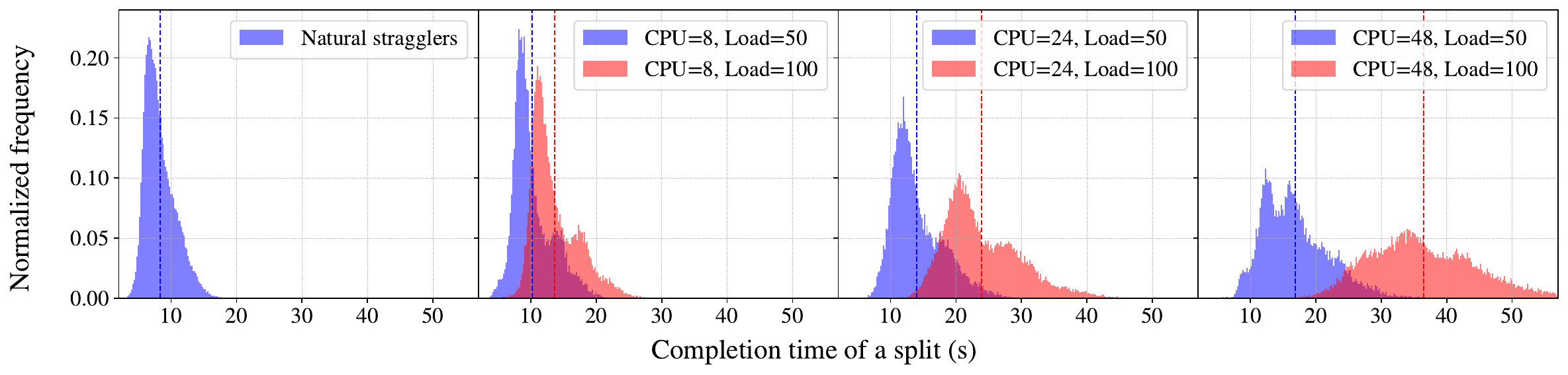}
\caption{Quantifying the effect of natural stragglers and \texttt{stress-ng} stragglers. Dashed vertical lines indicate the averages of histograms. See \secn{stragglerdetails} for details.}\label{fig:plot_hist_grid}
\end{figure*}

Given our quantification of the effect of inducing stragglers, we can induce a combination of stragglers with different \texttt{stress-ng} parameters to obtain a desired target straggler profile (e.g., to match a profile observed in a real system). For example, assume there exists $N$ histograms (i.e., \texttt{stress-ng} profiles) similar to those we obtained in \fig{plot_hist_grid}. Let $\bx_i\in\Rel^d$, $i\in\{1,\dots,N\}$, be the vector representation of a histogram, where $d$ is the number of histogram bins along the $x$-axis, and $\bx_i$ encodes $y$ values (normalized frequencies). In general we have $d\gg N$. We can approximate a given target straggler profile $\by\in\Rel^d$ by taking a convex combination of $\bx_i$ to minimize an error criterion. For example, one can approximate $\by\in\Rel^d$ with $\sum_{i=1}^{N}p_i\bx_i$ by solving the convex optimization problem
\begin{equation} \label{eqn:convexcomb}
\min_{p_1,\dots,p_N} \LnrmS{\by-\sum_{i=1}^{N}p_i\bx_i}
~~\text{s.t.}~~
p_i\geq0, \sum_{i=1}^{N}p_i=1.
\end{equation}
for a norm $\LnrmS{\cdot}$. Taking convex combinations of histograms amounts to combining stragglers stochastically. This means, per \eqn{convexcomb}, that we generate stragglers according to \texttt{stress-ng} parameters in $i$th histogram with probability $p_i$. 

\subsubsection{Inducing Temporal Stochasticity} \label{secn:expresultec2}
So far, we have outlined our approach to generating and combining stragglers. Next, our aim is to simulate a straggler environment that resembles a stochastic process, where the straggler profile changes randomly over time. To achieve this we employ the following two step scheme. First, we combine naturally occurring stragglers (left-most sub-figure in \fig{plot_hist_grid}) with \texttt{stress-ng} induced stragglers (one of the histograms in three right-most sub-figures in \fig{plot_hist_grid}). We employ random sampling from a Bernoulli($p$) distribution to determine whether or not to induce stragglers. Specifically, we induce stragglers using \texttt{stress-ng} with probability $p$, and do not induce stragglers with probability $1-p$. This sampling process is performed independently on each compute node. 

Second, every 60 seconds we repeat the process of random sampling and change the induced stragglers based on the result. Resampling randomizes the stressing behavior throughout the experiment. Resampling too often (e.g., every 2 seconds) would result in a homogeneous straggler behavior across workers. This would make it hard to effect time savings by replicating tasks. On the other hand resampling too infrequently (e.g., every 5 minutes) would make the straggler behavior persistent within the duration of the experiment. Setting the resample duration to about one minute strikes a good balance between these two extremes. 

\subsubsection{Straggler Profiles for Experimentation}
\begin{figure}
\reducebelowcaptionskip
\centering\includegraphics[width=1\columnwidth]{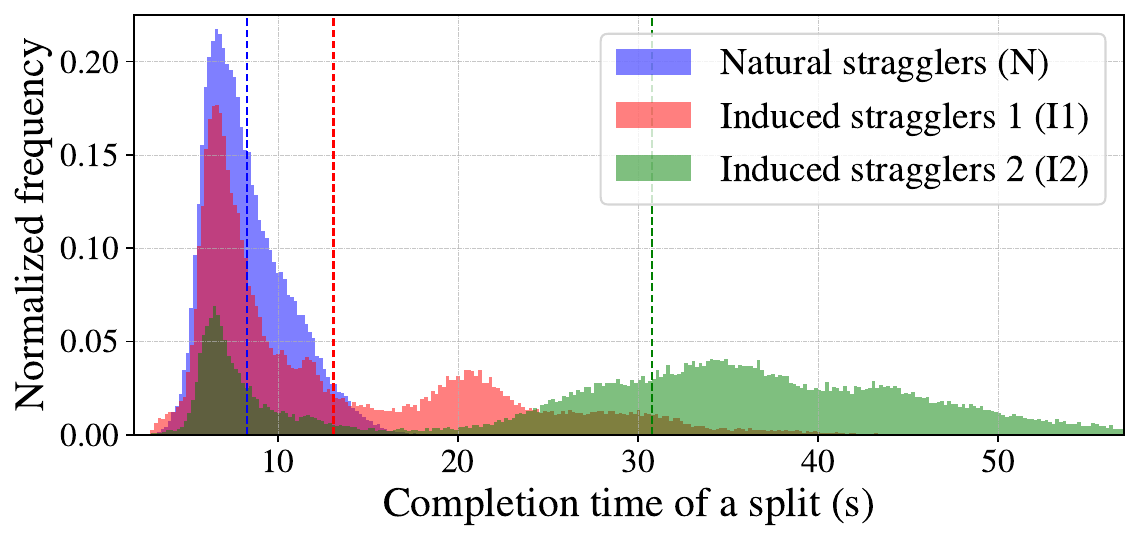}
\caption{Histograms obtained after stochastically combining natural stragglers and \texttt{stress-ng} stragglers. Dashed vertical lines indicate the averages. See \secn{expresultec2} for details.}
\label{fig:completiontimehist}
\end{figure}

We employ the techniques outlined so far to obtain three straggler profiles. \fig{completiontimehist} presents the histograms of task completion time corresponding to the three settings. In histogram (I1) we combine natural stragglers and \texttt{stress-ng} stragglers with parameters $\texttt{cpu}=24$ and $\texttt{load}=100$ (the third histogram in \fig{plot_hist_grid}). In this case we set $p=0.6$ in the Bernoulli distribution. 
In histogram (I2) we combine natural stragglers and \texttt{stress-ng} stragglers with parameters $\texttt{cpu}=48$ and $\texttt{load}=100$ (right-most histogram in \fig{plot_hist_grid}). In this case $p=0.95$ in the Bernoulli distribution. 
Histogram (N) corresponds to natural stragglers, i.e., without using the \texttt{stress-ng} stragglers. The PDFs we considered earlier in \fig{simstragglerpdf} in the simulated experiments were chosen to approximate the histograms in \fig{completiontimehist}. Specifically, we equalize the expected values of the PDFs in \fig{simstragglerpdf} to the averages of histograms in \fig{completiontimehist}. (To see this visually compare the vertical lines in the two figures.) 

\subsection{Completion Time Analysis}
In this section we measure the completion time of the proposed algorithm. We run the algorithm for different values of $\groupSize$ for the three different straggler settings presented in \fig{completiontimehist}. The results pertaining to the finishing time of the algorithm with Wikipedia-Uniform dataset are presented in \fig{yarntestresults} (i.e., back in \secn{backgroundamr}). Each point in \fig{yarntestresults} is obtained by averaging 20 repetitions. The labels of the plots indicate the straggler type. As one may expect, the plots obtained with induced stragglers are slightly above the plot obtained with natural stragglers. Although the three plots achieve their minimums at slightly different values of $\groupSize$, all plots follow a similar pattern when $\groupSize$ increases. As $\groupSize$ is increased from its initial value ($\groupSize=1$), computation time initially decreases and then after ($\groupSize\approx10\text{ to }15$) slightly increases. This is due to increased read/write operations in HDFS that take place over the communication network. In all three plots, setting $\groupSize=4$ is sufficient to realize most of the times savings. 

\begin{figure}
\reducebelowcaptionskip
\centering\includegraphics[width=0.995\columnwidth]{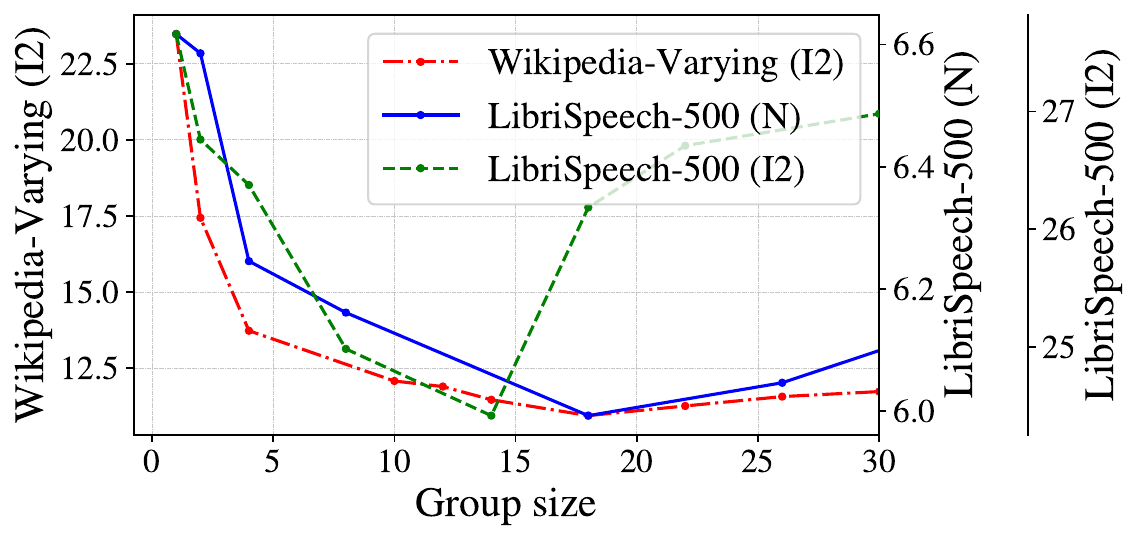}
\caption{\fig{yarntestresults}-equivalent for Wikipedia-Varying and LibriSpeech-500 datasets. Plots are obtained by averaging 10 repetitions. Three $y$-axis scales show the completion time in minutes. The legend indicates the dataset (as described in \secn{datasetsec2}) and the straggler profile (as presented in \fig{completiontimehist}).}
\label{fig:plot_line_yarnunequalwiki}
\end{figure}

\fig{plot_line_yarnunequalwiki} presents the results obtained using Wikipedia-Varying and LibriSpeech-500 datasets. In the three plots the observations are similar to those made with \fig{yarntestresults}. I.e., as $\groupSize$ increases, all plots follow a decrement of processing time in the beginning, followed by an increment of processing time. The key takeaway from \fig{plot_line_yarnunequalwiki} is that the proposed method works well even with varying task sizes. Out of all possible dataset-straggler profile combinations, we include only the combinations presented in \fig{plot_line_yarnunequalwiki} because of the limitations in our compute budget and the page limitations in the paper. In particular, we choose (I2) straggler profile because it has a higher straggler variability compared to (I1), thus it is easier to demonstrate the gains from our algorithm with (I2). For the LibriSpeech-500 dataset, we present results with natural stragglers as a baseline for that dataset.

\subsection{Split Assignment Visualization}
\begin{figure}
\centering\includegraphics[width=0.5\textwidth]{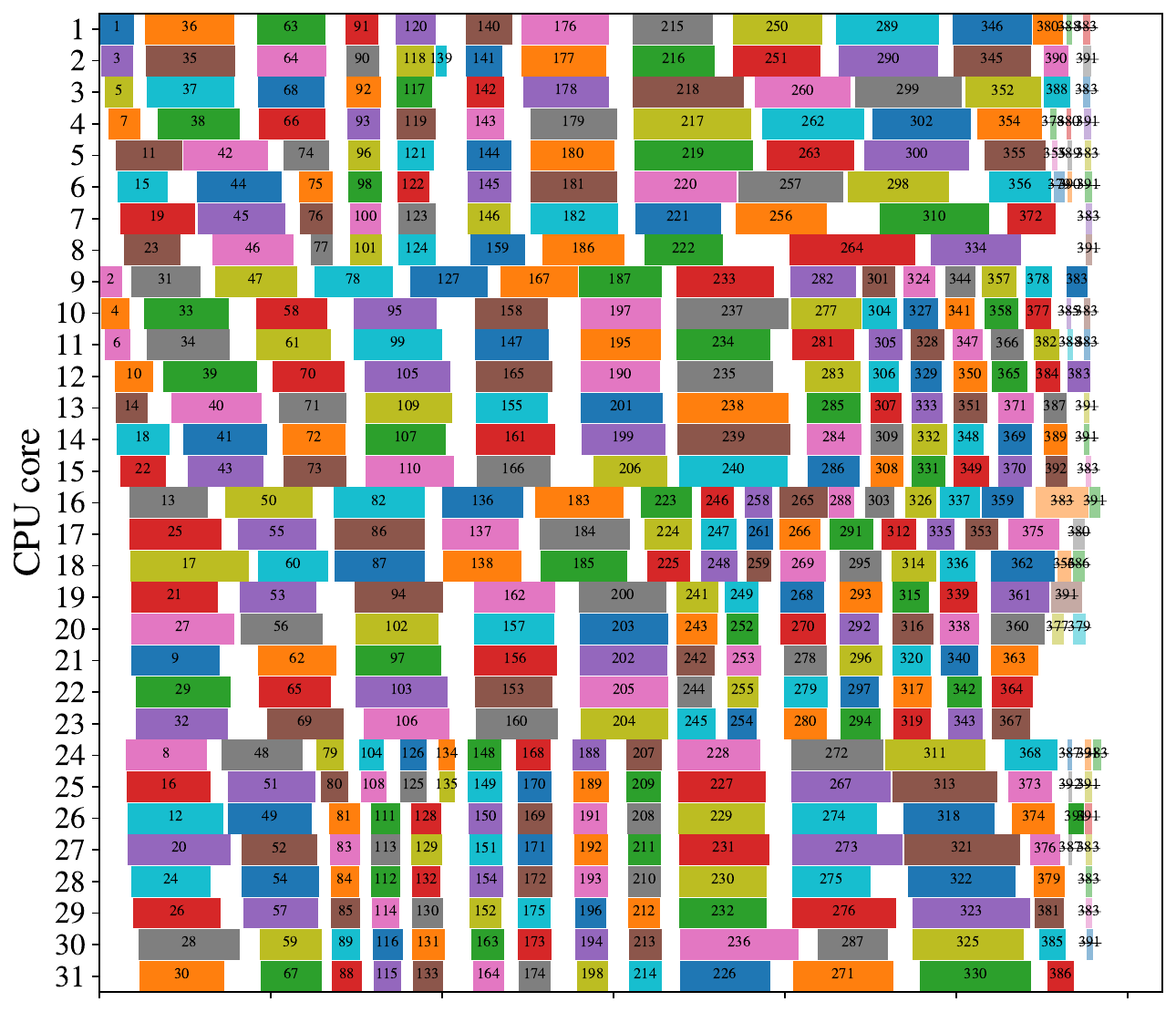}
\\ \vspace{-1.2ex}
\centering\includegraphics[width=0.5\textwidth]{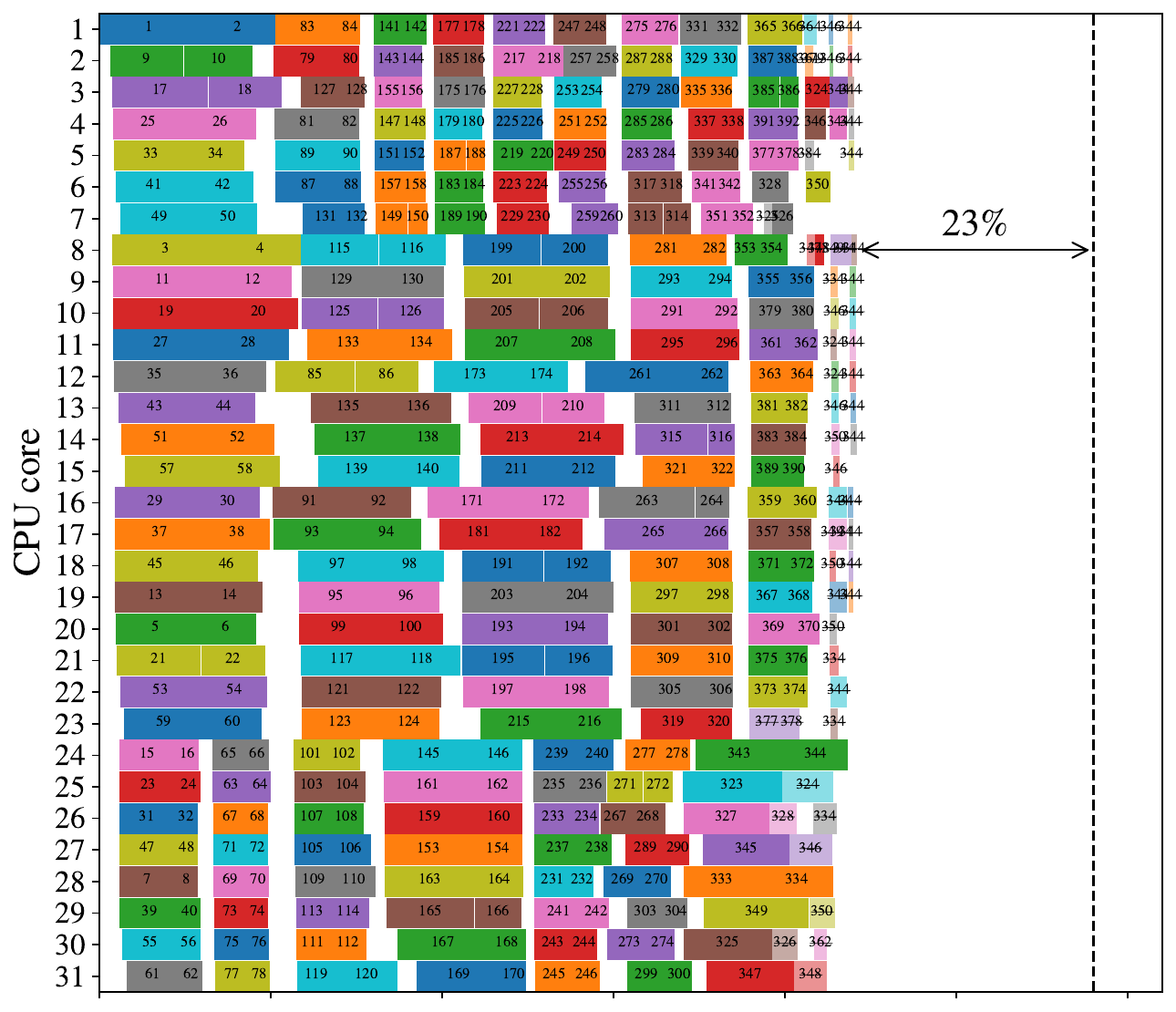}
\\ \vspace{-1.2ex}
\centering\includegraphics[width=0.5\textwidth]{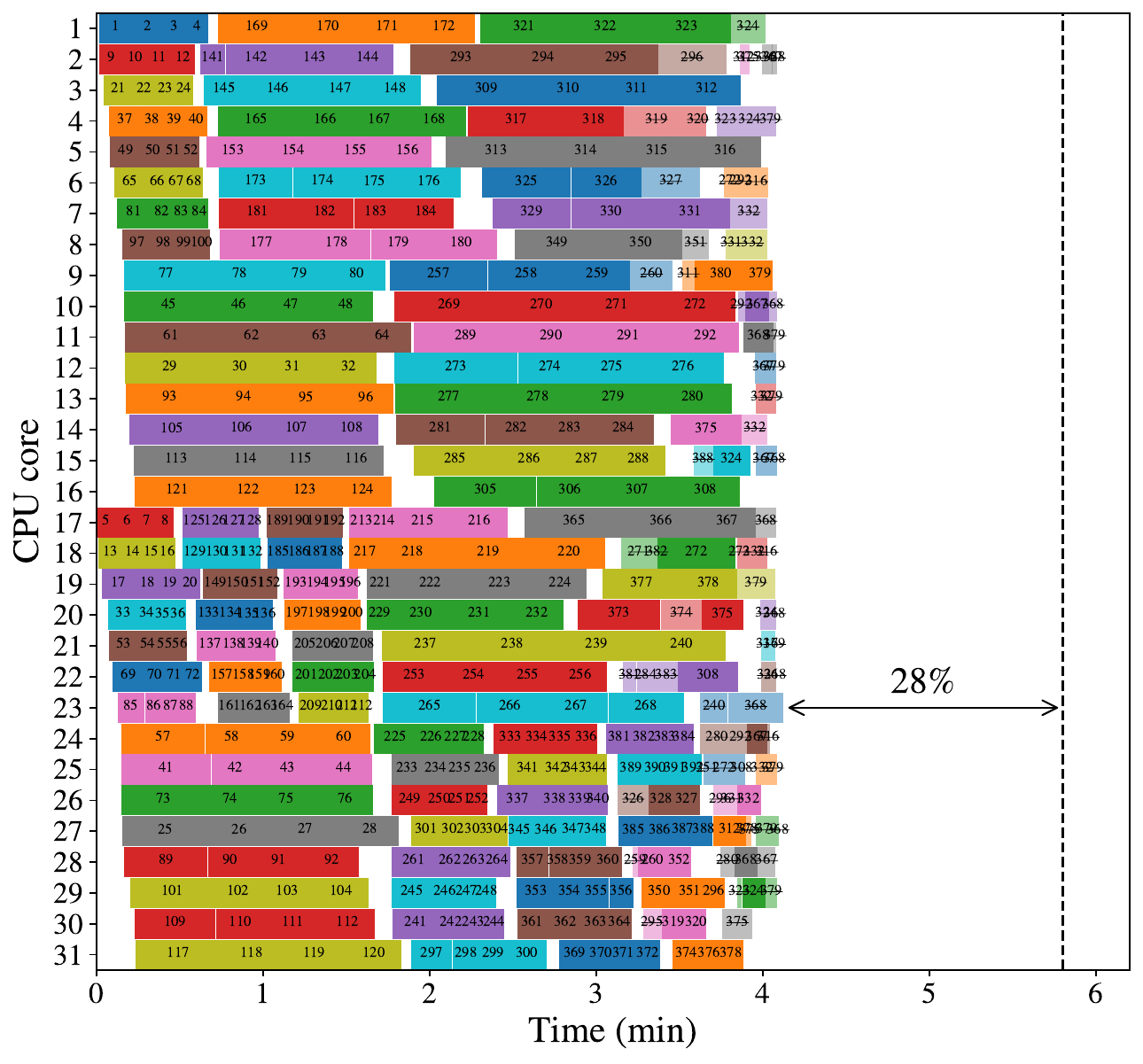}
\caption{Split assignments for $\groupSize=1$ (top), $\groupSize=2$ (middle) and $\groupSize=4$ (bottom) in Amazon EC2 experiments.}
\label{fig:amrvanillanorepec2}
\end{figure}

In this section we visualize how tasks may have been assigned to CPU cores while running experiments with the Wikipedia-Uniform dataset and when stragglers are induced according to (I1). \fig{amrvanillanorepec2} visualizes the task assignment for $\groupSize=1$, $\groupSize=2$ and $\groupSize=4$. Although \fig{amrvanillanorepec2} is generated from the traces we obtained from the EC2-based experiments, and is illustrative, we note that \fig{amrvanillanorepec2} may not be a 100\% accurate depiction of the actual task assignment on CPU cores. This is because in YARN it is not possible to track container assignments by the CPU core. We generate the visualizations from the task start/end time traces obtained from our experiments. We compare the start and end times of tasks and infer the assignment with the rule that tasks assigned to the same CPU core should not overlap. Therefore, the visualization generated may not be unique. 
For example, in the top sub-figure it is possible that CPU core 1 processes task 35 instead of task 36, and CPU core 2 processes task 36 instead of task 35. 

\fig{amrvanillanorepec2} helps visualize how the algorithm skips over some tasks. We observe that, due to straggling, some tasks take longer to process than others. For example, in the top sub-figure the processing of task 36 by CPU core 1 takes longer to process than task 1. The white space to the left of each task represents the processing overhead. The takeaway from \fig{amrvanillanorepec2} is that increasing $\groupSize$ yields a smaller completion time. 

